\documentclass[%
twocolumn,
superscriptaddress,
 amsmath,amssymb,
 aps,
 prx
]{revtex4-2}
\usepackage[english]{babel} %francais, polish, spanish, ...
\usepackage[T1]{fontenc}
\usepackage[ansinew]{inputenc}

\usepackage{amsmath}
\usepackage{amsthm}
\usepackage{amsfonts}
\usepackage{bbold}
\usepackage{xcolor}
\usepackage{soul}
\usepackage{dsfont}
\usepackage{MnSymbol}

\usepackage{graphicx}% Include figure files
\usepackage{dcolumn}% Align table columns on decimal point
\usepackage{bm}% bold math
\usepackage[colorlinks=true,allcolors=blue]{hyperref}% add hypertext capabilities
\usepackage{dirtytalk}
\usepackage{braket}
\usepackage{comment}
\usepackage{multirow}
\usepackage{makecell}
\usepackage{tabularx}
\usepackage{colortbl}
\usepackage{siunitx}

\usepackage{natbib}

\newcommand{\B}{\mathbf{B}}
\newcommand{\M}{\mathbf{M}}
\renewcommand{\H}{\mathbf{H}}
\newcommand{\Hd}{\mathbf{H}_\text{d}}
\newcommand{\Hdi}{\mathbf{H}_\text{d,in}}

\newcommand{\phii}{\phi_\text{d,in}}
\newcommand{\phio}{\phi_\text{d,out}}

\newcommand{\Rp}{R_\text{p}}

\newcommand{\Rch}{R_\text{c3}}
\newcommand{\bh}{b_3}
\newcommand{\Bh}{B_3}
\newcommand{\bq}{b_2}
\newcommand{\Bq}{B_2}
\newcommand{\bu}{b_1}
\newcommand{\Bu}{B_1}
\newcommand{\bo}{b_4}

\newcommand{\bd}{b_5}

\newcommand{\bdo}{b_6}

\newcommand{\Ih}{I_3}

\newcommand{\Nh}{N_3}
\newcommand{\Vp}{V_\text{p}}
\newcommand{\Sp}{S_\text{p}}
\newcommand{\diff}{\text{d}}

\newcommand{\Buo}{B^\text{ini}_{\text{1}}}
\newcommand{\Bqo}{B^\text{ini}_{\text{2}}}
\newcommand{\Bho}{B^\text{ini}_{\text{3}}}
\newcommand{\wzini}{\omega_0}
\newcommand{\eref}[1]{Eq.~(\ref{#1})}
\newcommand{\fref}[1]{Fig.~\ref{#1}}
\newcommand{\tref}[1]{Tab.~\ref{#1}}
\newcommand{\sref}[1]{Sec.~\ref{#1}}
\newcommand{\aref}[1]{App.~\ref{#1}}

\begin{document}

\title{Macroscopic Quantum Interference of the Center-of-Mass Motion of Levitated Superconducting Microparticles enabled by Magnetic Higher-Order Traps}
\author{Jaume Cunill-Subiranas}
\thanks{These two authors contributed equally to this work.}
\affiliation{Departament de F\'isica, Universitat Aut\`onoma de Barcelona, 08193 Bellaterra, Barcelona, Spain}
\author{Fabian Resare}
\thanks{These two authors contributed equally to this work.}
\affiliation{Department of Microtechnology and Nanoscience (MC2), Chalmers University of Technology, SE-412 96 Gothenburg, Sweden}
\author{Suocheng Zhao}
\affiliation{Institute for Quantum Materials and Technology, Karlsruhe Institute of Technology, 76131 Karlsruhe, Germany}
\author{Sofia Qvarfort}
\email{sofia.qvarfort@fysik.su.se}
\affiliation{Department of Physics, Stockholm University, SE-106 91 Stockholm, Sweden}
\affiliation{Nordita, KTH Royal Institute of Technology and Stockholm University, Hannes Alfv\'ens v\"ag 12, SE-114 19 Stockholm, Sweden}
\author{A. Metelmann}
\email{anja.metelmann@kit.edu}
\affiliation{Institute for Quantum Materials and Technology, Karlsruhe Institute of Technology, 76131 Karlsruhe, Germany}
\affiliation{Institute for Theory of Condensed Matter, Karlsruhe Institute of Technology, 76131 Karlsruhe, Germany}
\affiliation{Institut de Science et d'Ing\'enierie Supramol\'eculaires, UMR No. 7006, CNRS, University of Strasbourg, 67081 Strasbourg, France}
\author{Carles Navau}
\email{carles.navau@uab.cat}
\affiliation{Departament de F\'isica, Universitat Aut\`onoma de Barcelona, 08193 Bellaterra, Barcelona, Spain}
\author{Witlef Wieczorek}
\email{witlef.wieczorek@chalmers.se}
\affiliation{Department of Microtechnology and Nanoscience (MC2), Chalmers University of Technology, SE-412 96 Gothenburg, Sweden}

%%%%%%%%%%%%%%%%%%%%%%%%%%%%%%
%%%%%%%%%%%%%%%%%%%%%%%%%%%%%%

\begin{abstract}
We show how magnetostatic higher-order multipole traps can be used to generate macroscopic quantum interference of the motion of levitated superconducting microparticles. An appropriate combination of multipolar magnetic fields offers great versatility in constructing various trap potentials, including anharmonic trap potentials such as Duffing or double-well types. Crucially, the anharmonic trap potentials realize a nonlinearity on the order of hundred times the zero-point motion, i.e., on a length scale below nanometers. These anharmonic potentials allow for the generation of quantum features of the center-of-mass motion of a magnetically levitated superconducting microparticle. Importantly, they can be easily generated with a static arrangement of coils, requiring only that the current running through them is tunable. We propose protocols exploiting the versatility of the magnetic trap landscape to generate non-Gaussian motional states. We solve the dynamics of the center-of-mass motion of the particle in phase space and analyze its parameter dependence. Furthermore, we give a recipe to distinguish classical from quantum behavior in a statistically meaningful way through measurement of the position of the particle. Our results open a path to accessing the quantum regime of the center-of-mass motion of objects with masses larger than picogram, i.e., $10^{13}$ atomic mass units. This will enable fundamental physics experiments for studying the transition between quantum and classical behavior, exploring the intersection between quantum physics and gravity as well as probing of certain types of dark matter.
\end{abstract}
%%%%%%%%%%%%%%%%%%%%%%%%%%%%%%
%%%%%%%%%%%%%%%%%%%%%%%%%%%%%%

\maketitle

%%%%%%%%%%%%%%%%%%%%%%%%%%%%%%
%%%%%%%%%%%%%%%%%%%%%%%%%%%%%%
\section{Introduction} 
%%%%%%%%%%%%%%%%%%%%%%%%%%%%%%
%%%%%%%%%%%%%%%%%%%%%%%%%%%%%%

The generation of nonclassical states of macroscopic objects is of fundamental relevance. It enables exploring the validity of quantum mechanics for larger objects \cite{Arndt1999,leggettTestingLimitsQuantum2002,schrinskiMacroscopicQuantumTest2023, Pedalino26}, for testing models for wavefunction collapse \cite{diosiUniversalMasterEquation1987,penroseGravityRoleQuantum1996,collapse_RevModPhys}, or for probing the interface between quantum mechanics and gravity \cite{ boseSpinEntanglementWitness2017,marlettoGravitationallyInducedEntanglement2017,Aspel2018,Bose25}. Furthermore, nonclassical states of macroscopic objects can also be used in novel quantum sensing schemes, see, e.g., \cite{Johnsson2016,weissLargeQuantumDelocalization2021,coscoEnhancedForceSensitivity2021,omahenUltracoldMechanicalQuantum2026}, which specifically exploit the mass of the object.

Levitation of nano- or micrometer-sized particles constitutes a suitable platform for generating nonclassical motional states of massive objects \cite{millenQuantumExperimentsMicroscale2020,gonzalez-ballestero_2021} and can be based on optical \cite{Romero-Isart_2010,chang_2010,barker_2010,delicCoolingLevitatedNanoparticle2020,tebbenjohannsQuantumControlNanoparticle2021,ranfagniTwodimensionalQuantumMotion2022,kambaQuantumSqueezingLevitated2025}, electrical \cite{Goldwater_2019,dania_2024}, or magnetic traps \cite{PRL_Romero-Isart2012,Cirio12,waarde2016thesis,slezakCoolingMotionSilica2018,vinanteUltralowMechanicalDamping2020,Gutierrez2023,Hofer2023}. In particular, magnetic levitation of superconductors offers a promising alternative to optical or electrical traps for macroscopic quantum experiments as it provides a noise-free trap environment that does not pose a mass limit on the levitated object \cite{PRL_Romero-Isart2012, Cirio12, QST_Pino2018}. Magnetic quadrupole traps---such as those generated by anti-Helmholtz coils---were proposed as an approach to trap superconducting microparticles \cite{PRL_Romero-Isart2012} and have recently been implemented at millikelvin temperatures in both on-chip  \cite{Gutierrez2023} and macroscopic traps \cite{Hofer2023,SchmidtPRApplied}. However, these systems are limited to harmonic confinement for small motional amplitudes and thus require additional means to prepare nonclassical states.
%offer no control over the shape of the potential.
Methods to generate nonclassical motional states of trapped particles can be based on using projective measurements \cite{romero-isartLargeQuantumSuperpositions2011,Romero-Isart2011,QST_Pino2018}, external \cite{PRL_Romero-Isart2012,Cirio12,Johnsson2016,ramannairMassiveQuantumSuperpositions2025a} or internal coupling \cite{albrechtTestingQuantumGravity2014,hoangElectronSpinControl2016,boseSpinEntanglementWitness2017,marlettoGravitationallyInducedEntanglement2017,delordSpincoolingMotionTrapped2020,marshmanConstructingNanoobjectQuantum2022} to qubits, or the use of non-linear potentials \cite{QST_Pino2018,rakhubovskyStroboscopicHighorderNonlinearity2021,Borah21,PRL_Roda-Llordes2024,Neumeier2024,PRXQuantum.5.030312,Casulleras24,Grochowski2025}. The latter method is appealing as it allows a direct dynamic manipulation of the center-of-mass motion of the particle without requiring external means.

In our work, we propose and analyze a protocol that exploits dynamical state evolution in an anharmonic potential \cite{PRL_Roda-Llordes2024,rodallordes2023numerical,Neumeier2024} with the aim to generate a non-Gaussian state of the center-of-mass motion of a trapped superconducting microsphere. Our proposal brings together (i) dynamic state evolution in anharmonic potentials that realize a nonlinearity on a length scale of hundred times the zero-point motion (i.e., sub-nanometer), (ii) a recipe to generate the required magnetic fields with static coil arrangements, and (iii) stationary trapping of the same microparticle allowing for repeated experiments to acquire sufficient statistics. Furthermore, our proposal is implemented within the platform of superconducting magnetic levitation that combines the desirable built-in features \cite{PRL_Romero-Isart2012,Cirio12,Johnsson2016,QST_Pino2018} of (iv) operation at cryogenic vacuum conditions, (v) use of noiseless magnetic trap fields, both reducing decoherence, and (vi) the capability to couple the particle's motion to superconducting quantum circuits for efficient read-out \cite{PRL_Romero-Isart2012,Cirio12,Johnsson2016,QST_Pino2018}. We will show that these features allow us to realize and verify a non-Gaussian state of a micrometer-sized superconducting particle with a total extension of about 100\,picometer.

Our work presents a toolbox based on higher-order magnetic traps (HOTs), in which the dominant terms in the magnetic potential arise from multipoles higher than the quadrupole---such as the hexapole and beyond. While higher-order multipole contributions have previously been observed as perturbations to harmonic magnetic traps for large motional amplitudes \cite{Gutierrez2023}, HOTs realize these as the principal trap terms, i.e., for amplitudes on the order of hundred times the ground state motion, i.e., about 0.2\,nm. HOTs enable flexible, anharmonic trap confinement and provide the anharmonicity required to generate nonclassical states of motion. We focus specifically on spherical-hexapole traps \cite{PRA_Bergeman1987,PRA_Weinstein1995}--the lowest-order HOTs--that offer both strong confinement and high symmetry. When the hexapolar field is combined with homogeneous or quadrupolar fields, they enable a rich set of trap landscapes, including double-well and other anharmonic potentials. We calculate the dynamics of the motion of the microparticle in such potentials and identify parameter regimes, which allow to distinguish quantum from classical signatures in a statistically meaningful sense. 

In the following, we describe the general recipe to obtain various magnetic potentials for trapping of superconducting microspheres (\sref{sec:HOTS}), derive the accompanying Hamiltonians (\sref{sec:HAM}), and analyze the motional dynamics numerically in phase space (\sref{sec:dynamics}). We illustrate the generation of nonclassical states of motion by using the Wigner function and discuss how to distinguish between quantum and classical dynamics and the parameter-dependence of the protocol (\sref{sec:protocols}), assisted by analyzing the dynamics through an approximate model analytically (\sref{sec:analytics}). Finally, we elaborate on the challenges in an experimental implementation (\sref{sec:implementation}) and give a summary and outlook of our work (\sref{sec:theend}). We include an appendix, which gives details for the derivation of HOTs, the numerical analysis in phase space, the analytic modeling, and additional results.

\begin{figure}[t!hbp]
    \centering
    \includegraphics[width=1.0\linewidth]{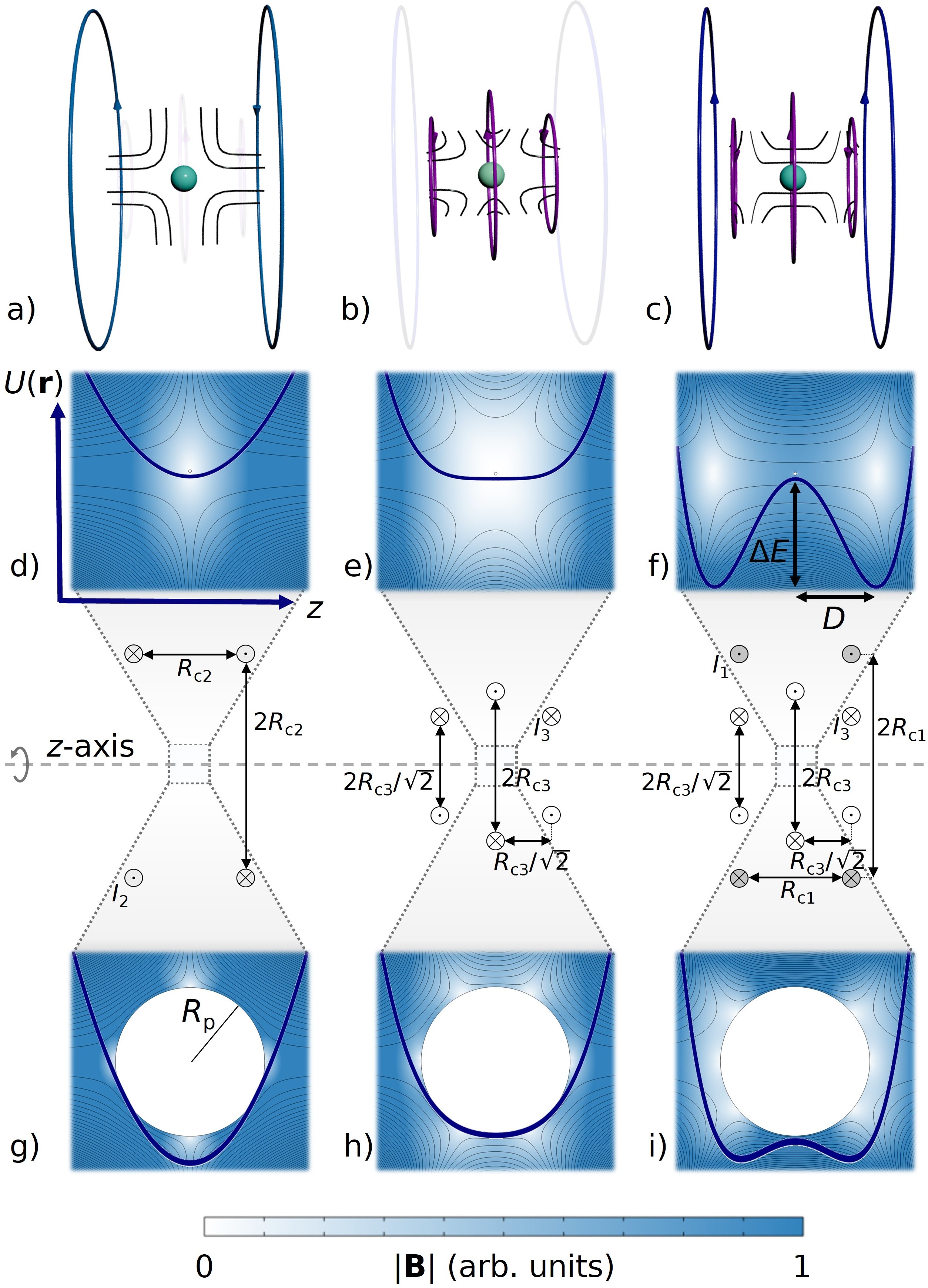}
    \caption{Harmonic and anharmonic magnetic traps for superconducting particles. Coil configurations, magnetic fields, and mechanical potentials for a magnetic quadrupole (a,d,g), magnetic hexapole (b,e,h), and double-well potential obtained from the superposition of a uniform magnetic field and magnetic hexapole (c,f,i). These fields are induced in the central region of a suitable arrangement of coils, all of them with cylindrical symmetry in the $z$-direction. The crosses and dots indicate the sense of the current in the coils. Equal magnitudes of currents are represented by the same colours. The magnetic field modulus is shown for small particles $\Rp\ll r_\text{rms}$ (d-f) and large particles $\Rp\gtrsim r_\text{rms}$ (g-i), calculated via finite-element simulations. The mechanical potential along the $z$-direction is shown in thick blue lines [see \eref{eq:H_z}]. 
    }
    \label{fig:coils_and_potentials}
\end{figure}

%%%%%%%%%%%%%%%%%%%%%%%%%%%%%%
%%%%%%%%%%%%%%%%%%%%%%%%%%%%%%
\section{Magnetic higher-order trapping fields}\label{sec:HOTS}
%%%%%%%%%%%%%%%%%%%%%%%%%%%%%%
%%%%%%%%%%%%%%%%%%%%%%%%%%%%%%

We consider trapping of a spherical particle of radius $\Rp$ that is a perfect diamagnet with magnetic susceptibility $\chi_m=-1$, i.e., a type-I superconductor with $\Rp\gg \lambda_L$, where $\lambda_L$ is the London penetration depth of the utilized superconductor. For the elemental type-I superconductors being used in experiments \cite{waarde2016thesis,latorreChipBasedSuperconductingMagnetic2022,Gutierrez2023}, this holds as $\Rp$ is in the micrometer scale and $\lambda_L$ is typically on the order of $50\,$nm. The mechanical potential experienced by such a particle under any source-free applied magnetostatic induction field, $\B_\text{a}$, is given by
\begin{equation}\label{eq:full_pot}
    \begin{split}
    \mathcal{U}(\textbf{r})=&\frac{1}{2\mu_0}\int_{\Vp}|\B_\text{a}(\textbf{r}+\textbf{r}')|^2\diff V'\\&+\frac{\mu_0}{2}\sum_{n=0}^{\infty}(n+1)\Rp^{-(2n+1)}\sum_{m=-n}^{n}|a_{nm}(\textbf{r})|^2,
\end{split}
\end{equation}
where $\textbf{r}$ is the position of the center of the particle, $\textbf{r}'$ denotes the position relative to the center of the particle, and $\Vp=4\pi\Rp^3/3$ is its volume. The coefficients $a_{nm}$ are given by the applied field evaluated on the surface of the particle, $\Sp=4\pi\Rp^2$, as
\begin{equation}\label{eq:coefficients}
    \resizebox{0.48\textwidth}{!}{$\displaystyle
    a_{nm}(\mathbf{r}) = 
    -\frac{\Rp^{n+2}}{\mu_0(n+1)} \oint_{\Sp}
    \left[\B_\text{a}(\mathbf{r} + \Rp\mathbf{e}_{r'}) \cdot \mathbf{e}_{r'}\right]
    {Y_n^m}^*(\theta', \varphi') \, \text{d}\Omega',
    $}
\end{equation}
where $Y_n^{m*}(\theta',\varphi')$ are the conjugated orthonormalized spherical harmonics and $\text{d}\Omega'=\sin\theta'\text{d}\theta'\text{d}\varphi'$ is the solid angle differential in spherical coordinates. A detailed derivation of these equations, which enables calculating the potential landscape of a trapped superconducting particle in a straightforward way is found in \aref{app:magnetic_energy}.

If the \textit{applied} field---the field in the absence of the particle---is uniform, or can be approximated as such, over the volume occupied by the particle, then the potential simplifies to $\mathcal{U}(\textbf{r})=\frac{\pi\Rp^3}{\mu_0}|\B_\text{a}(\bf{r})|^2$. We refer to this limit as the small-sphere approximation \cite{PRL_Romero-Isart2012,prat-camps2016,QST_Pino2018,bort-soldevila2024}. In this case, the range of admissible mechanical potentials is only constrained by Maxwell's source-free magnetostatic equations $\nabla\cdot\B_\text{a}=0$ and $\nabla\times\B_\text{a}=0$ \cite{PQE_Wing1984,PRA_Bergeman1987}. Hence, one could create any mechanical potential $\mathcal{U}\propto\nabla\cdot(\phi\nabla \phi)$, with $\phi$ being any magnetic scalar potential, $\B_\text{a}=-\mu_0\nabla \phi$.

For larger particles and arbitrary applied fields---where the full \eref{eq:full_pot} must be used---the coefficients $a_{nm}$ are generally difficult to obtain analytically. However, many of these coefficients vanish for some fields with particular symmetries, such as axial symmetry. This situation arises specifically for fields that can be decomposed into simple sums of pure cylindrical-multipole components, which can be generated by certain configurations of coaxial current loops. Then, a configuration of $w$ coaxial current loops or dipoles is required to generate the $2w$-multipolar term,  where $w$ denotes the order of the field \cite{PRA_Bergeman1987,PRA_Weinstein1995}. In such cases, as we state in \aref{app:beyond_hexa}, the highest order of the multipolar fields determines which coefficients are non-zero. The coefficients for the uniform field, the spherical quadrupole [\fref{fig:coils_and_potentials}(d)], and for the first HOT field, the spherical hexapole [\fref{fig:coils_and_potentials}(e)], are explicitly given in \aref{app:fields_and_coeff}. This first HOT field can be induced in the center of a spherical coil configuration consisting of three coaxial coils, see \fref{fig:coils_and_potentials}(b) \cite{PRA_Bergeman1987,PRA_Weinstein1995}. The field strength from the center of this hexapolar trap increases with the distance squared, unlike the quadrupole, which increases linearly. This feature makes the hexapole a candidate for building anharmonic traps. 

In the following, using shifted coordinates $\tilde{\textbf{r}}=(\tilde{x},\tilde{y},\tilde{z})\equiv\textbf{r}+\textbf{r}'$, we consider the combination of fields up to hexapolar order that leads to the total applied field as
\begin{equation}\label{eq:applfield}
    \begin{split}
    \B_\text{a}(\tilde{\textbf{r}})=&-\left[\bh (\tilde{z}\tilde{x})+\bq\frac{\tilde{x}}{2}\right]\mathbf{e}_{x}-\left[\bh (\tilde{z}\tilde{y})+\bq\frac{\tilde{y}}{2}\right]\mathbf{e}_{y}+\\&\left[\bh\frac{2\tilde{z}^2-\tilde{x}^2-\tilde{y}^2}{2}+\bq\tilde{z}+\bu\right]\mathbf{e}_{z}\, .
    \end{split}
\end{equation}
This applied field depends on the parameters $\bh$, the strength of the hexapole in T/m$^2$ [from  \eref{eq:hexapole_field} and \eref{eq:Bh}, see \fref{fig:coils_and_potentials}(b)]; $\bq$, the strength of the quadrupole in T/m [from \eref{eq:quadrupole_field} and \eref{eq:B2}, see \fref{fig:coils_and_potentials}(a)]; and $\bu$, the magnetic induction field in T for the uniform field [from  \eref{eq:uniform_field} and \eref{eq:Helmholtz_coils_strength}]. 

These coefficients can be either positive or negative depending on the direction of the currents. Importantly, obtaining a symmetric field profile with two minima along the $z$ direction---required to generate a symmetric double-well potential---necessitates $\bh$ and $\bu$ to have opposite signs and $\bq=0$. This situation is shown in \fref{fig:coils_and_potentials}(f). Note that the total magnetic field strength that can be applied will ultimately be constrained by the critical field $B_\text{c}$ of the utilized superconductor. Thus, $\sum_w|B_w|\leq B_\text{c}$ with 
\begin{equation}\label{eq:bigb}
B_w\equiv b_w\Rp^{w-1},
\end{equation}
and all $B$ have units of T.

%%%%%%%%%%%%%%%%%%%%%%%%%%%%%%
%%%%%%%%%%%%%%%%%%%%%%%%%%%%%%
\section{Hamiltonian for Anharmonic Magnetic Trap Landscapes}\label{sec:HAM}
%%%%%%%%%%%%%%%%%%%%%%%%%%%%%%
%%%%%%%%%%%%%%%%%%%%%%%%%%%%%%

%%%%%%%%%%%%%%%%%%%%%%%%%%%%%%
\subsection{General Hamiltonian}
%%%%%%%%%%%%%%%%%%%%%%%%%%%%%%

To evaluate the particle's dynamics, we construct the Hamiltonian
\begin{align}\label{eq:classical_hamiltonian}
\mathcal{H}&=\mathcal{T}+\mathcal{U},
\end{align}
for a superconducting particle placed in the applied field given by \eref{eq:applfield}. Considering only translational degrees of freedom, the kinetic term is given by $\mathcal{T}=\textbf{p}^2/2m$, where $m$ is the mass and $\textbf{p}$ the linear momentum of the particle. The potential $\mathcal{U}$ is obtained by calculating the coefficients $a_{nm}$ for the field of \eref{eq:applfield}, which are given in  \eref{eq:coeff_uniform}, \eref{eq:coeff_quadrupole}, and \eref{eq:coeff_hexapole}. We obtain
\begin{equation} \label{eq:complete_hamiltonian}
    \begin{split}
        \mathcal{U}=\frac{\pi \Rp^3}{36\mu_0}\bigg[&36\bu^2-36\bh\bu(x^2+y^2-2z^2)\\ &+24\bq(2\bh\Rp^2+3\bu+3\bh z^2)z\\&+3\bq^2\big[4\Rp^2+3(x^2+y^2+4z^2)\big]\\&+\bh^2\big[6\Rp^4+16\Rp^2(x^2+y^2+3z^2)\\&+9(x^4+2x^2y^2+y^4+4z^4)\big]\bigg]\,.
    \end{split}
\end{equation}
The potential $\mathcal{U}$ contains terms up to fourth order in position, and the motion of the particle along the $z$-direction is decoupled from the motion along the $x$ and $y$ directions, while the latter two are coupled. Note that the decoupling of $z$ is possible only for field combinations up to hexapolar order. In HOTs with contributions beyond the hexapole, the axial and transverse degrees of freedom generally become coupled.

The fact that the motion along the $z$ direction decouples from the $x$ and $y$ directions allows us to constrain our analysis to the $z$ direction only. Then, the potential along the $z$ direction is given by
\begin{align} \label{eq:H_z}
        \nonumber \mathcal{U}_{z} & =\frac{\pi\Rp^3}{6\mu_0}\bigg[\left(6\bh^2\right)z^4+\left(12\bh\bq\right)z^3\\
        \nonumber & +\left(12\bh\bu+6\bq^2+8\Rp^2\bh^2\right)z^2\\
        \nonumber &+\left(12\bq\bu+8\Rp^2\bh\bq\right)z\\
        &+6\bu^2+2\Rp^2\bq^2+\Rp^4\bh^2\bigg].
\end{align}
It is instructive to analyze how the particle's radius influences its expected dynamics. To this end, we distinguish 'small' and 'large' particles. A particle is considered small---in the sense of the small-sphere approximation introduced earlier---if its radius is much smaller than its motional amplitude, i.e., $\Rp \ll z_\text{rms}$. Large particles correspond to the case when $\Rp \gtrsim z_\text{rms}$. When a pure hexapolar field is applied, i.e., $\bu=\bq=0$ and $\bh\neq0$, a small particle would be exposed to a dominantly anharmonic potential proportional to $z^4$ [see \fref{fig:coils_and_potentials}(e)]. In contrast, a large particle would see a harmonic potential [see \fref{fig:coils_and_potentials}(h)]. 

We assume now for the purposes of state generation that an initial motional state of the particle is prepared through ground-state cooling in a harmonic potential. In the case of \eref{eq:H_z}, this harmonic potential is given by the term proportional to $z^2$, which can be obtained either by using a pure quadrupolar field (i.e., $\bu=\bh=0$),  or by using a pure hexapolar field in the large particle approximation, which will be automatically realized through sufficient pre-cooling of the particle's motional amplitude. In this harmonic potential, the \textit{initial} harmonic trap frequency of the particle is
\begin{equation}
    \wzini=\sqrt{\frac{6\Bho\Buo+3(\Bqo)^2+4(\Bho)^2}{2\mu_0\varrho\Rp^2}}\, ,
\end{equation}
where $\varrho=3m/(4\pi\Rp^3)$ is the mass density of the particle, with an associated zero-point motional amplitude of $z_{\text{zpf}}=\sqrt{\hbar/(2m\wzini)}$. After state purification, the fields required for the particle to explore a new potential are rapidly set to the desired values. We may write down a quantized Hamiltonian of the particle motion in the new potential \textit{in terms of the ladder operators of the initial harmonic trap} using the particle position $\hat{z}=z_{\text{zpf}}(\hat{a}^\dagger+\hat{a})$ and its momentum $\hat{p}=p_{\text{zpf}}(\hat{a}-\hat{a}^\dagger)/i$ with $p_{\text{zpf}}=\sqrt{\hbar m \wzini/2}$ and $[\hat{a},\hat{a}^\dagger]=1$. We obtain (neglecting constant terms)
\begin{align}\label{eq:H_quantum_complete}
    \hat{\mathcal{H}}_{z}&=-\frac{\hbar\wzini}{4}(\hat{a}-\hat{a}^\dagger)^2+\hbar\sum_{j=1}^{4}\Omega_j{(\hat{a}+\hat{a}^\dagger)}^j,
\end{align}
with rates $\Omega_j$
\begin{eqnarray}\label{eq:rates_short}
        \nonumber\Omega_1 &=&c_0\left(6\Bq\Bu+4\Bh\Bq\right)z_{\text{zpf}}/\Rp\, ,\\
        \nonumber\Omega_2&=&c_0\left(6\Bh\Bu+3\Bq^2+4\Bh^2\right)z_{\text{zpf}}^2/\Rp^2\, ,\\
        \nonumber\Omega_3&=&c_0\left(6\Bh\Bq\right)z_{\text{zpf}}^3/\Rp^3\, ,\\
        \Omega_4&=&c_0(3\Bh^2)z_{\text{zpf}}^4/\Rp^4\, ,
\end{eqnarray}
where $c_0=\pi\Rp^3/(3\mu_0\hbar)$. We will use this Hamiltonian to analyze protocols exploiting the nonlinear dynamics of the particle's motion. Note that the system of equations for the rates $\Omega_j$ in terms of fields $B_w$ is overdetermined, as further discussed in \aref{app:overdetermined_system}.

In the following, we focus on two types of nonlinear potentials that contain terms up to order four in the operators, which lead to the generation of non-Gaussian motional states: a double-well potential (denoted as DWP) and a pure Duffing potential (denoted as DP). In general, the flexibility of magnetic traps extends beyond these specific choices. Alternative configurations, such as a tilted double-well potential or other anharmonic potentials, might also be accessed through tailored coil arrangements. Interested readers may use \tref{tab:magnetic_HOT_coeff} to, for example, construct potentials from even higher-order fields.

%%%%%%%%%%%%%%%%%%%%%%%%%%%%%%
\subsection{Double-well potential (DWP)}
%%%%%%%%%%%%%%%%%%%%%%%%%%%%%%

We can create a narrow double-well potential by introducing a field detuning $d\equiv(3\Bu+2\Bh)/5(3\Bu-2\Bh)$ between the ratio of $B_1$ and $B_3$ and setting $\Bq=0$. We obtain

\begin{align}\label{eq:H_quantum_completeDP}
    \hat{\mathcal{H}}_{z}^{\text{DWP}}&=-\frac{\hbar\wzini}{4}(\hat{a}-\hat{a}^\dagger)^2+\hbar\Omega_2{(\hat{a}+\hat{a}^\dagger)}^2+\hbar\Omega_4{(\hat{a}+\hat{a}^\dagger)}^4,
\end{align}
with $\Omega_2=-40c_0\Bh^2z_{\text{zpf}}^2d/(1-5d)\Rp^2$ and $\Omega_4=3c_0\Bh^2z_{\text{zpf}}^4/\Rp^4$. Note that $d$ needs to be non-zero to have the harmonic term with non-zero $\Omega_2$ in the Hamiltonian. The distance $D_\text{DWP}$ between the lobes of the double-well is
\begin{equation}\label{eq:lobe_distance}
    \frac{D_{\text{DWP}}}{z_{\text{zpf}}}=\pm \frac{\Rp}{z_{\text{zpf}}}\sqrt{-\frac{\Bu}{\Bh}-\frac{2}{3}}\, ,
\end{equation}
with $\Bu/\Bh=-2(1+5d)/3(1-5d)$. The height of the barrier separating the two lobes is 
\begin{equation}\label{eq:height_double-well}
    \frac{|\Delta E|}{\hbar\wzini}=\frac{\pi\Rp^3}{\hbar\mu_0\wzini}\left(\Bu+\frac{2}{3}\Bh\right)^2.
\end{equation}
When the particle is initially in the center of the double-well, $|\Delta E|$ will be the expectation value of the potential energy imparted to the particle. 

The length scale of the double-well $D_{\text{DWP}}$ is set only via the ratio of the magnetic field strength. For example, to obtain a narrow double-well potential, the ratio between the fields $\Bu$ and $\Bh$ should only very slightly detuned from $-2/3$, see \eref{eq:lobe_distance}. Importantly, $D_{\text{DWP}}$ is not limited by use of a specific geometry or light of a certain wavelength, such as in optical levitation \cite{PRL_Roda-Llordes2024,Neumeier2024}. With HOTs, anharmonic potentials are instead generated through the combination of precise currents  and the field mixing obtained from the potential being $\mathcal{U}\propto{|\mathbf{B}_{\text{a}}|}^2$. Thus, we can realize relevant nonlinearities on the order of the zero-point motion \cite{samantaNonlinearNanomechanicalResonators2023}. 

Note that a particle could be initialized in the DWP either on the top of the barrier, or in either of the wells. The latter situation is similar to the one considered in Ref.~\cite{PRL_Roda-Llordes2024}, and we call this displaced DWP (D-DWP).

%%%%%%%%%%%%%%%%%%%%%%%%%%%%%%
\subsection{Duffing potential (DP)}
%%%%%%%%%%%%%%%%%%%%%%%%%%%%%%

A pure Duffing potential, i.e., a pure $\hat{z}^4$ quartic potential, can be obtained by setting the detuning $d=0$, such that $\Bh=-3\Bu/2$ and $\Bq=0$. We get
\begin{align}\label{eq:H_quantum_completeKP}
    \hat{\mathcal{H}}_{z}^{\text{DP}}&=-\frac{\hbar\wzini}{4}(\hat{a}-\hat{a}^\dagger)^2+\hbar\Omega_4{(\hat{a}+\hat{a}^\dagger)}^4,
\end{align}
with $\Omega_4$ from \eref{eq:rates_short}.

\begin{figure*}[t!hbp]
    \centering
    \includegraphics[width=1.0\linewidth]{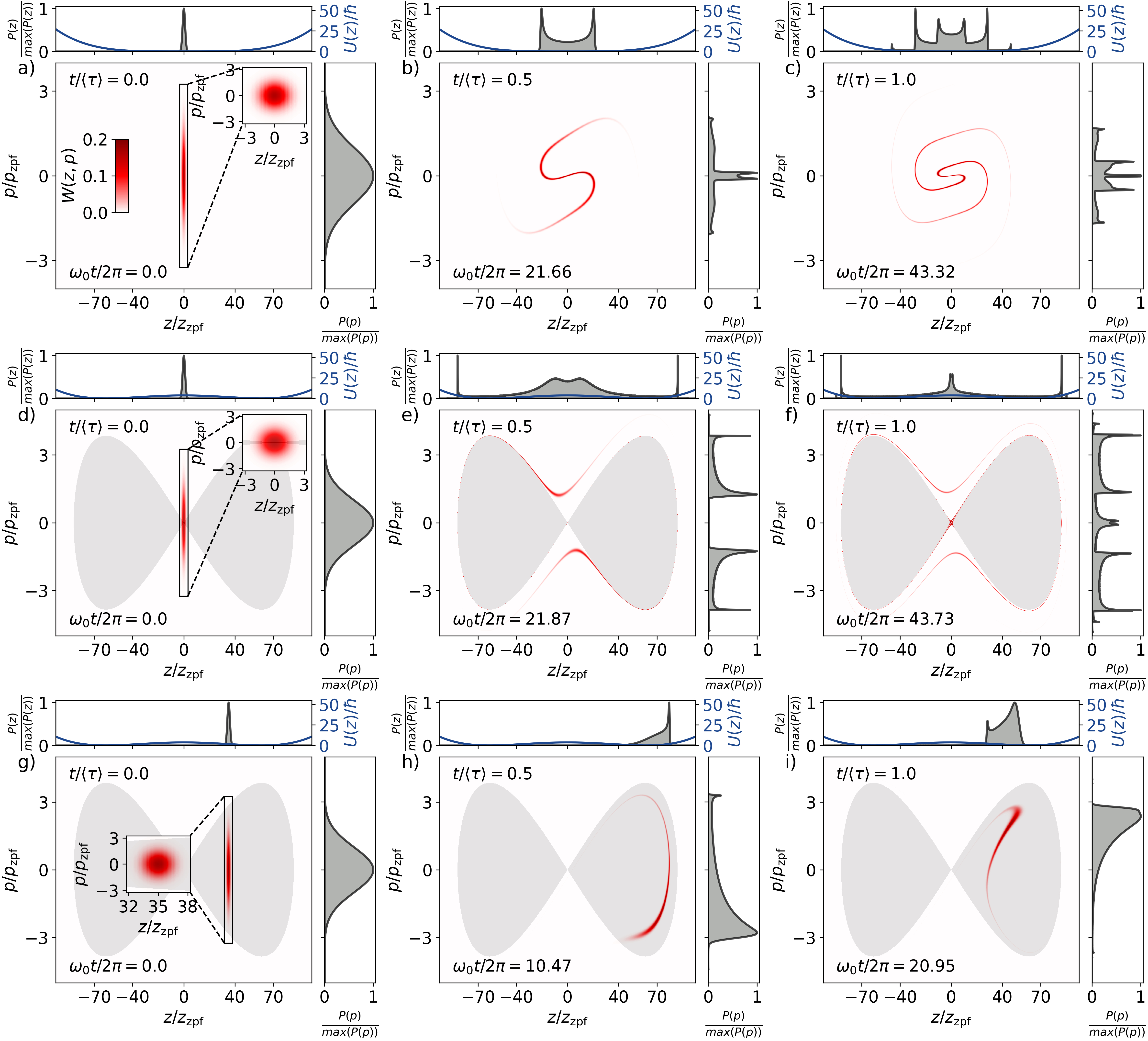}
    \caption{Classical dynamics of the motion of a 1\,\textmu m radius particle. We show Wigner functions and their marginal probability distributions in position (top panels) and momentum (side panels) at different times, given in terms of the original harmonic oscillator time scale. State evolution in  (a,b,c) the DP potential (parameters: $\clubsuit$ in \tref{tab:parameters}), (d,e,f) the DWP potential (parameters: $\spadesuit$ in \tref{tab:parameters}), and (g,h,i) the D-DWP potential for an initial state displaced by $35z_{\text{zpf}}$ (parameters: $\spadesuit$ in \tref{tab:parameters}). In the case of the DWP and D-DWP potentials, the region of $\mathfrak{m}>1$ according to condition \eref{eq:m_geq_1} is shaded, where the motion is constrained to one half of the double well. Note that the initial state shown in panels (a,d,e) is the ground state (see zoom-in).}
    \label{fig:classical_dynamics}
\end{figure*}

%%%%%%%%%%%%%%%%%%%%%%%%%%%%%%
%%%%%%%%%%%%%%%%%%%%%%%%%%%%%%
\section{Phase-space dynamics}\label{sec:dynamics}
%%%%%%%%%%%%%%%%%%%%%%%%%%%%%%
%%%%%%%%%%%%%%%%%%%%%%%%%%%%%%

%%%%%%%%%%%%%%%%%%%%%%%%%%%%%%
\subsection{Classical case}\label{sec:dynamics_classical}
%%%%%%%%%%%%%%%%%%%%%%%%%%%%%%

We now consider the dynamics of the particle's motion when exposed to the nonlinear potentials. To this end, we use a phase-space representation of the particle's motion to study its time evolution. This allows us to evaluate the Wigner function $W(z,p,t)$, which captures the complete information of the particle's motional state. This method is preferred in cases where a density matrix formulation is impractical due to a large state delocalization, which is the case in our protocols. We thereby follow the approach outlined in Ref.~\cite{rodallordes2023numerical}. 

We consider the Hamiltonian from \eref{eq:H_quantum_completeDP} with rates $\Omega_2$ and $\Omega_4$ and use from now on unitless quadratures ($z/z_{\text{zpf}}\rightarrow z$, $p/p_{\text{zpf}}\rightarrow p$)

\begin{equation}\label{eq:x4_hamilton}
    \mathcal{H}/\hbar=\frac{\omega_0}{4}p^2+\Omega_2z^2+\Omega_4z^4.
\end{equation}

The conservative classical dynamics of the Wigner function, denoted as $W_{\text{c}}$, can be found by taking the Poisson bracket between $W_{\text{c}}$ and $\mathcal{H}/\hbar$, yielding
\begin{equation} \label{eq:klein_kramers_double_well}
    \resizebox{0.48\textwidth}{!}{$\displaystyle
    \mathcal{L}_\text{c} W_{\text{c}}\equiv\frac{\partial W_{\text{c}}(z,p,t)}{\partial t}=-\frac{\omega_0p}{2}\frac{\partial W_{\text{c}}}{\partial z}+\left(2\Omega_2z+4\Omega_4z^3\right)\frac{\partial W_{\text{c}}}{\partial p},
    $}
\end{equation}
where we defined $\mathcal{L}_\text{c}$ as a Liouville operator responsible for the classical dynamics. The resulting \eref{eq:klein_kramers_double_well} is a Klein-Kramers equation, and its solution can be obtained through Liouville's theorem (see \aref{app:liouville_classical}). We obtain the exact solution for the classical trajectories of the motion (see also Refs.~\cite{RieraCampeny2024analytical, salas2014}), denoted as $z_\text{c}$ and $p_\text{c}$,  

\begin{align}\label{eq:classical_solutions}
    \nonumber z_\text{c}(t)&=c_1\text{cn}\left(\nu t+c_2, \mathfrak{m}\right),\\
    p_\text{c}(t)&=-2\frac{\nu}{\omega_0} c_1\text{sn}\left(\nu t+c_2, \mathfrak{m}\right)\text{dn}\left(\nu t+c_2, \mathfrak{m}\right),
\end{align}
where $\nu=\sqrt{\omega_0(\Omega_2+2\Omega_4c_1^2)}$, $\mathfrak{m}=\omega_0\Omega_4c_1^2/\nu^2$ is the elliptic modulus, and the cn, sn, and dn-functions denote the respective Jacobi elliptic functions. In the case that $\Omega_4=0$, they reduce respectively to the cosine, sine, and identity functions, as would be the solution for a harmonic oscillator. Details on obtaining the integration constants $c_1$ and $c_2$ from the initial conditions can be found in \aref{app:int_constants_and_derivatives}. From the periodicity of the Jacobi elliptic functions, the solutions always exhibit a periodicity of $\tau=2\text{Re}(K(\mathfrak{m}))/\nu$ for the cases we consider (see \aref{app:int_constants_and_derivatives}), which can be used to obtain an approximate timescale for the dynamics. 

\fref{fig:classical_dynamics} shows the solutions of the classical dynamics for a 1\,\textmu m radius particle in the DP, DWP, and D-DWP protocols. In all protocols, the dynamics are propagated for one mean period time $\braket{\tau}$ of the initial state. For all cases, the state expands strongly along the position quadrature, while it remains at roughly the same extent along the momentum quadrature. The latter is due to the fact that we change the potential term in the protocol, but the momentum term remains unchanged. 

In the DP, once the state has expanded to a size large enough for the nonlinearity to become active, the component of the state with large displacement rotates faster and generates the spiraling structure visible in \fref{fig:classical_dynamics}(b) and \fref{fig:classical_dynamics}(c), see also Ref.~\cite{rodallordes2023numerical}. The DWP protocol generates a vastly different phase space structure with fast acceleration towards the turning point of the double well [\fref{fig:classical_dynamics}(e)] and later on a partial revival towards the center [\fref{fig:classical_dynamics}(f)]. The trajectory in the D-DWP protocol [\fref{fig:classical_dynamics}(g,h,i)] is restricted to one lobe of the double well, i.e., within the shaded region marking the case of the elliptic modulus being $\mathfrak{m}>1$. For initial conditions starting within this region, which is present only in the D-DWP, the motion is constrained to the lobe that they start in.

As the dynamics in all of these cases are classical, the Wigner functions are always positive. However, in the marginal probability distributions peaks can be observed, as, e.g., observable in the side panels of \fref{fig:classical_dynamics}(b,c,e,f), which resemble wave interference.

%%%%%%%%%%%%%%%%%%%%%%%%%%%%%%
\subsection{Quantum case}\label{sec:dynamics_full}
%%%%%%%%%%%%%%%%%%%%%%%%%%%%%%

To model the full dynamics, we include a Liouvillian term producing quantum dynamics given by
\begin{equation}
    \mathcal{L}_\text{q}=-8\Omega_4z\frac{\partial^3}{\partial p^3},
\end{equation}
which is found by specializing the general expression \cite{gardiner-zoller2004} to potentials up to order four in the coordinates. Furthermore, noise processes during the dynamics are captured by the dissipative Liouvillian \cite{risken1985,rodallordes2023numerical}
\begin{equation}\label{eq:dissipativeLiouvillian}
    \mathcal{L}_\text{n}=\gamma_0\left(1+p\frac{\partial}{\partial p}\right)+\frac{2\gamma_0 k_{\text{B}}T}{\hbar\omega_0}\frac{\partial^2}{\partial p^2},
\end{equation}
and additional momentum diffusion processes could be taken into account by \cite{rodallordes2023numerical, Romero-Isart2011}
\begin{equation}\label{eq:excess_liouvillian}
    \mathcal{L}_\text{e}=\Lambda\frac{\partial^2}{\partial p^2},
\end{equation}
with $\Lambda$ a localization parameter \cite{gardiner-zoller2004,Romero-Isart2011}. Details for these Liouvillian terms are found in \aref{app:liouvillians}. The mean occupation of the environment $\bar{n}$ at $\omega_0$ is related to the temperature $T$ in the high-temperature limit via $\bar{n}\approx k_{\text{B}}T/\hbar\omega_0$, which is about $2\cdot10^6$ at $T=10\,$mK with $\omega_0=2\pi\cdot100$ Hz. The nature of the limiting loss and noise processes in magnetic levitation experiments of superconducting microparticles at ultralow temperatures are still unknown. They could be related to, for example, external vibrations, magnetic field fluctuations, magnetic flux vortices, gas collisions, or black-body radiation, see, e.g.~Refs.~\cite{romero-isartLargeQuantumSuperpositions2011,PRL_Romero-Isart2012,Johnsson2016,QST_Pino2018,narasimhamoorthyMagneticNoiseMacroscopic2025,schutExpressionDecoherenceRate2025} for a detailed analysis. In our work, we assume for simplicity that coherence is limited by coupling to a thermal environment according to \eref{eq:dissipativeLiouvillian}. We therefore parametrize dissipation in terms of the quality factor $Q=\omega_0/\gamma_0$. Note that when the frictional term in \eref{eq:dissipativeLiouvillian} can be neglected \cite{PRXQuantum.5.030312}, one can use the mapping $\Lambda=\gamma_0\bar{n}/2$ to evaluate these different noise processes. Theoretical estimates predict quality factors for magnetically levitated superconducting microparticles in excess of $10^{10}$ \cite{PRL_Romero-Isart2012,Cirio12}, with current experiments at $2\cdot10^{7}$ \cite{Hofer2023}. Below, we show that a quality factor of at least $10^{11}$ is required to observe quantum features in the Wigner function, which is in the range of theoretically estimated values \cite{PRL_Romero-Isart2012,Cirio12}.

We make use of a rotating frame transformation with respect to the classical dynamics (Liouville frame, Ref.~\cite{rodallordes2023numerical}), in which we write the transformed Wigner function as $\Tilde{W}$. The main advantage of this procedure is that the phase space grid deforms along the classical trajectories, accurately capturing narrow regions in phase space without needing excessively fine discretization. The equation governing the time evolution in this frame is then
\begin{equation}\label{eq:liouville_frame_dynamics}
    \frac{\partial \Tilde{W}}{\partial t}=e^{-\mathcal{L}_\text{c} t}\left(\mathcal{L}_\text{n}+\mathcal{L}_\text{e}+\mathcal{L}_\text{q}\right)e^{\mathcal{L}_\text{c} t}\Tilde{W}.
\end{equation}
This equation now captures the full dynamics of the system in the Liouville frame, including an environment ($\mathcal{L}_{\text{n}}$, $\mathcal{L}_{\text{e}}$) and quantum effects ($\mathcal{L}_{\text{q}}$). We numerically obtain the solution to \eref{eq:liouville_frame_dynamics} by employing the computational method of Ref.~\cite{rodallordes2023numerical}, which we modified by performing the classical trajectory part of the simulation using the exact solution, i.e., \eref{eq:classical_solutions}. We elaborate further on these modifications in \aref{app:jacobi_eval}. Note that all resulting Wigner functions that we show are given in the original frame.

\begin{figure*}[t!hbp]
     \centering
     \includegraphics[width=\linewidth]{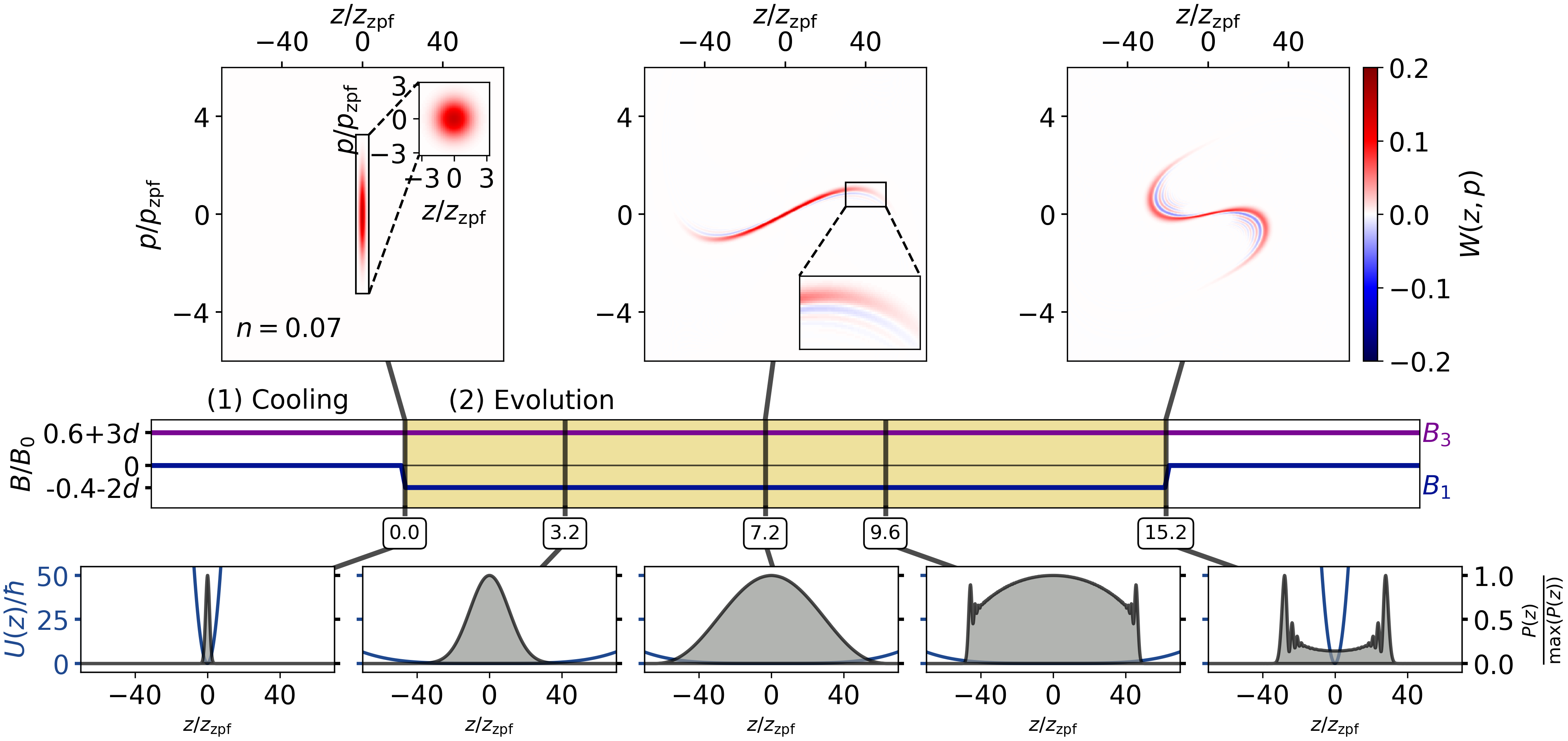}
     \caption{Non-Gaussian motional state of a levitated superconducting microparticle. DP case (parameters: $\clubsuit$ in \tref{tab:parameters}): The motion is initially near ground-state cooled ($n=0.07$) in a harmonic potential, including very weak coupling to the environment ($Q=10^{12}$). At time $t=0$, the $z^4$-potential is abruptly switched on, while maintaining $\Bh=\Bho$. The upper panel shows the evolution of the Wigner function, the middle panel shows the switching of coils to realize certain magnetic fields, and the lowermost panel shows the position probability distribution $P(z)$ as well as the potential $\mathcal{U}_{z}$. We observe that the $z$-motion first experiences motional squeezing, followed by a generation of interference fringes upon surpassing the turning point. At the end of the protocol the position of the particle is measured. The particle could be recaptured in a harmonic potential for this purpose.}
\label{fig:Wigner_protocols}
\end{figure*}

%%%%%%%%%%%%%%%%%%%%%%%%%%%%%%
%%%%%%%%%%%%%%%%%%%%%%%%%%%%%%
\section{Non-Gaussian motional states}\label{sec:protocols}
%%%%%%%%%%%%%%%%%%%%%%%%%%%%%%
%%%%%%%%%%%%%%%%%%%%%%%%%%%%%%

In the following, we will analyze the generation of non-Gaussian states of motion of a trapped superconducting microparticle with a particle radius $\Rp=1$\,\textmu m in the nonlinear magnetic trap potentials we introduced. Parameters for the magnetic field to arrive at the desired potentials are given in \aref{app:table_data}.

%%%%%%%%%%%%%%%%%%%%%%%%%%%%%%
\subsection{Dynamics of the Wigner function}
%%%%%%%%%%%%%%%%%%%%%%%%%%%%%%
 
We exemplarily focus on the DP protocol with parameters $\clubsuit$ from \tref{tab:parameters}, and show the time evolution of the Wigner function in that potential in \fref{fig:Wigner_protocols}. We observe that the particle's motion experiences first a squeezing action, i.e., coherent inflation \cite{romero-isartCoherentInflationLarge2017,QST_Pino2018,PRL_Roda-Llordes2024}, until a point where the motional amplitude has become large enough for the quartic part of the potential to become active. Then, fringes appear first in the momentum distribution. At this point in time, a possibility could be to retrap the particle in a harmonic potential, let the particle's motion evolve dynamically until the fringes map to position \cite{QST_Pino2018}, and then measure the position quadrature. However, we find that higher visibility fringes are directly generated in the position distribution once the motional turning points in the DP have been surpassed at times $\omega_0t/2\pi>9$. Furthermore, the extent of the fringes in position is significantly larger than the ones in momentum. In principle, the evolution in the anharmonic potential could continue for longer times. Yet, as the experiment is anticipated to be decoherence limited, the motional state should be measured as soon as visible fringes appear. When the initial phonon occupation is slightly increased from $n=0.07$ to $n=0.5$, the fringes lose visibility, see \fref{fig:Wigner_DP_App} in \aref{app:add_results}. Importantly, we can identify that a motional center-of-mass state with an extent of about 80 times the zero-point motion, i.e., 0.1\,nm, can be created with HOTs, accounting for residual phonon occupation and a finite quality factor.

We also show the Wigner function dynamics for a $\Rp=1$\,\textmu m in the DWP protocol in \fref{fig:Wigner_DWP_App} in \aref{app:add_results}. Also in the case of the DWP, we observe negative parts with fringes in the Wigner function evidencing nonclassicality. We note that cases similar to the D-DWP protocol have been treated previously in Refs.~\cite{PRL_Roda-Llordes2024, RieraCampeny2024analytical}.

\begin{figure}[t!hbp]
    \centering
    \includegraphics[width=\linewidth]{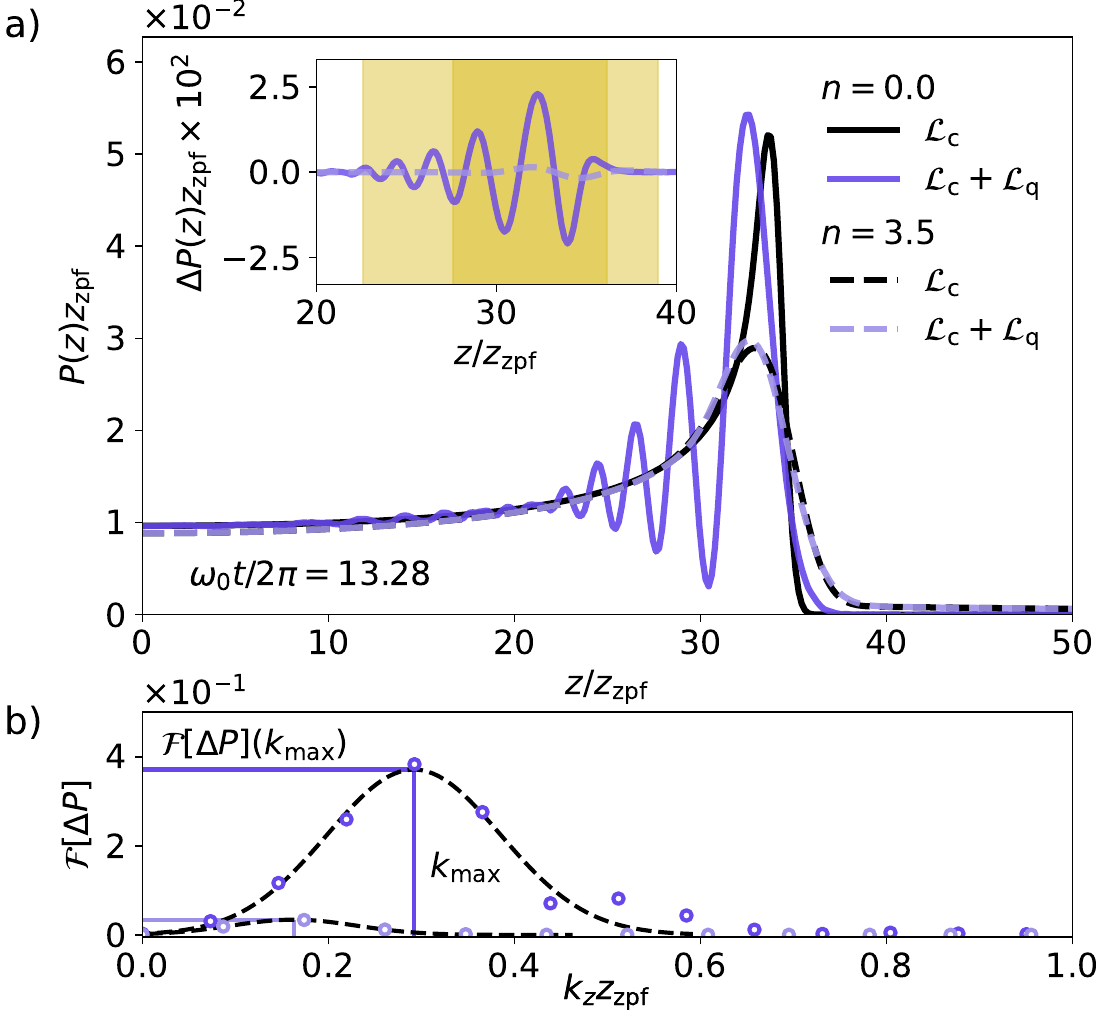}
    \caption{Discerning quantum from classical dynamics in the DP protocol. (a) Probability distribution of the position quadrature for an initial phonon occupation of $n=0$ (dark purple, straight) and $n=3.5$ (light purple, dashed), compared to the classical dynamics (black). For these simulations, $Q=\infty$. The inset shows the difference between the quantum and classical probability distributions $\Delta P$. (b) Fourier transformation of $\Delta P$, restricted to the respective highlighted areas in the inset of (a). Gaussian fits (dashed lines) are used to extract the peak wavenumber $k_{\text{max}}$ and maximal amplitude $\mathcal{F}[\Delta P](k_{\text{max}})$. }
    \label{fig:transform}
\end{figure}

%%%%%%%%%%%%%%%%%%%%%%%%%%%%%%
\subsection{Comparison of quantum and classical dynamics}
%%%%%%%%%%%%%%%%%%%%%%%%%%%%%%

Importantly, we now analyze the ability of the experiment to distinguish quantum from classical dynamics. To this end, we stop the evolution in the DP at a time when visible fringes in the position probability distribution $P(z)$ have appeared, see  \fref{fig:transform}(a). We observe that the quantum dynamics produce a clearly visible fringe pattern in the position distribution, while such a pattern is absent for the classical dynamics. To quantify this observation, we plot the difference $\Delta P$ between the two probability distributions in the inset of \fref{fig:transform}(a) and transform it into reciprocal space to $\mathcal{F}[\Delta P]$, which is shown in \fref{fig:transform}(b). We observe a clear peak with an amplitude $\mathcal{F}[\Delta P](k_{\text{max}})$ at peak wavenumber $k_{\text{max}}$. In the following analysis, we will use the amplitude $\mathcal{F}[\Delta P](k_{\text{max}})$ and wavenumber $k_{\text{max}}$ of this peak as a measure for the experimental accessibility for verifying quantum dynamics with the protocol.

%%%%%%%%%%%%%%%%%%%%%%%%%%%%%%
\subsection{Finite phonon number}
%%%%%%%%%%%%%%%%%%%%%%%%%%%%%%

We consider the case of non-zero phonon occupation of the initial state that will be exposed to the nonlinear potential by using the initial Wigner function 
\begin{equation}\label{eq:W_initial}
    W(z,p,0)=\frac{1}{2\pi\sigma_z\sigma_p}e^{-\frac{1}{2}\left[\left(z/\sigma_z\right)^2+\left(p/\sigma_p\right)^2\right]}
\end{equation}
with $n=\braket{\hat{a}^\dagger \hat{a}}=\frac{1}{4}(\sigma_z^2+\sigma_p^2)-\frac{1}{2}$. The $\sigma_z$ and $\sigma_p$ terms account for the initial uncertainty of the state in the position and momentum quadratures, respectively. We restrict our analysis to states with a symmetric initial uncertainty $\sigma_z=\sigma_p=\sqrt{2n+1}$, but note that squeezed initial states could be analyzed as well and may be beneficial for the protocol \cite{PRXQuantum.5.030312}. 

\fref{fig:transform}(a) shows the position probability distribution for $n=3.5$, corresponding to an initial state purity of $\mathcal{P}=1/(2n+1)=12.5\%$ \cite{parisPurityGaussianStates2003}. When considering the difference between the quantum and classical dynamics [inset of \fref{fig:transform}(a)], fringes are still visible, but with reduced amplitude compared to the $n=0$ case. \fref{fig:transform}(b) shows this behavior clearly as a reduction of the peak height in reciprocal space, simultaneously occurring at a smaller wavenumber. However, quantum features persist even in this impure occupation case, similar to recent experiments observing noisy cat states \cite{Yang2025}.

\begin{figure}[t!hbp]
    \centering
    \includegraphics[width=\linewidth]{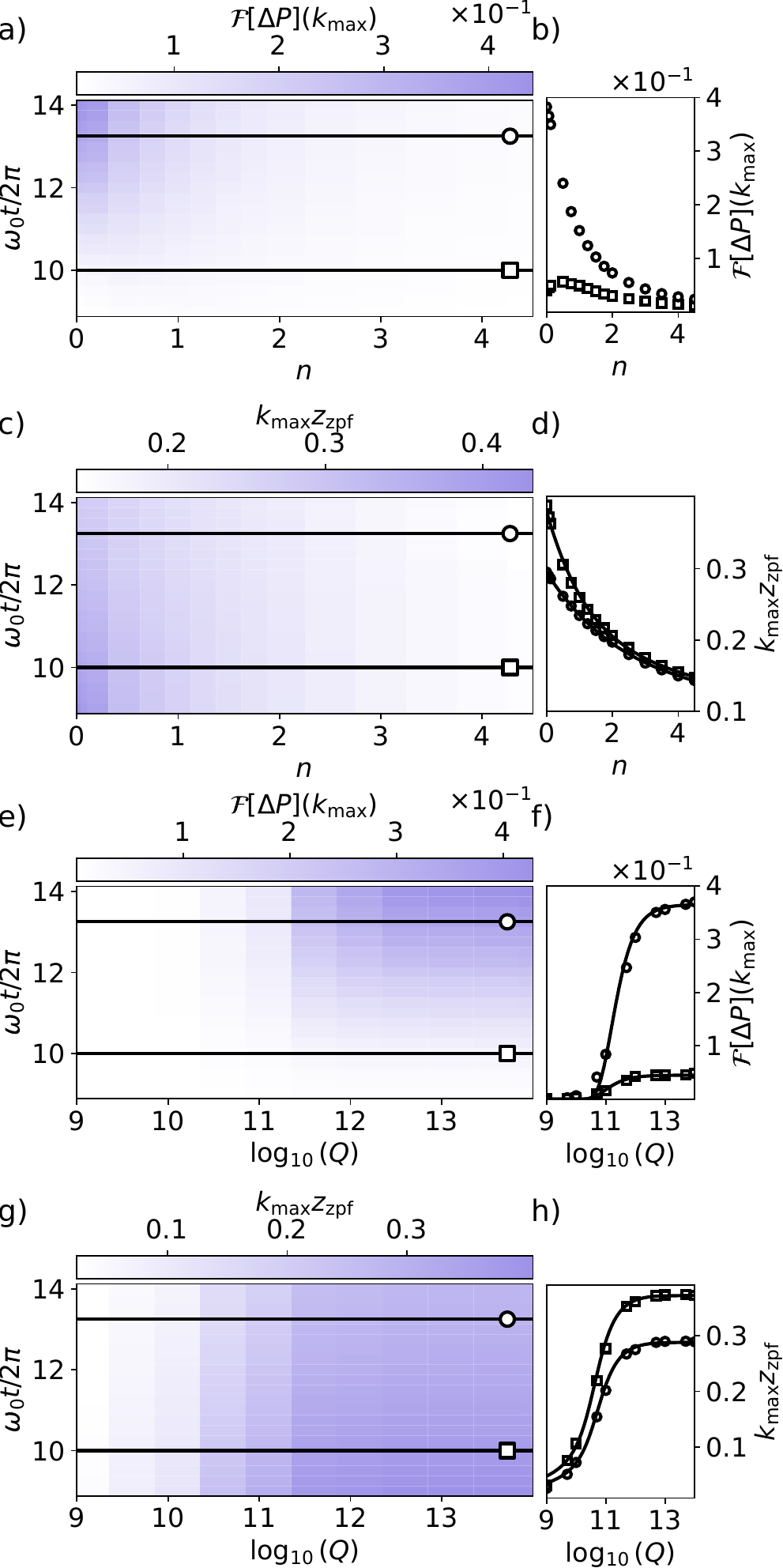}
    \caption{Observability of quantum dynamics in the DP protocol. We show in panels (a, e) the fringe amplitude $\mathcal{F}[\Delta P](k_{\max})$ and in (c, g) the peak wavenumber $k_{\max}$. Panels (a, c) show these quantities for different initial phonon occupations $n$, while panels (e, g) show them dependent on the quality factor $Q$. Panels (b, f) show line cuts of the fringe amplitude and panels (d, h) of the wavenumber at times $\omega_0t/(2\pi)=\{10, 13.28\}$. The data points are results from numerical simulation of the dynamics, while the lines are fits to the approximate analytical model.}
    \label{fig:feasibility}
\end{figure}

\fref{fig:feasibility}(a) and \fref{fig:feasibility}(c) show a systematic decrease of the fringe amplitude and wavenumber, respectively, with the initial phonon occupation. Interestingly, for short protocol times, i.e., around $\omega_0t/2\pi\sim 10$, we observe that a non-zero phonon number results in the largest fringe amplitude, see \fref{fig:feasibility}(b). For longer protocol times, the maximum fringe amplitude increases and shifts towards smaller phonon numbers. This behavior can be understood by considering the mean period time $\braket{\tau}$ in the DP potential and its dependence on the initial conditions. A larger initial phonon number results in a faster timescale for the dynamics such that the nonlinearity of the DP can impact the state more strongly for the same evolution time, yielding an earlier fringe generation. Overall, we find that protocol DP is relatively robust against imperfect ground state cooling. Furthermore, a non-zero initial phonon occupation may even be advantageous.

%%%%%%%%%%%%%%%%%%%%%%%%%%%%%%
\subsection{Thermal decoherence}
%%%%%%%%%%%%%%%%%%%%%%%%%%%%%%

We now study the influence of the thermal environment on the experiment. To this end, we use the dissipative Liouvillian $\mathcal{L}_n$ to model thermal decoherence, assuming coupling of the particle's motion to a thermal environment at a temperature $T=10$\,mK with a dissipation rate $\gamma_0$, which is parametrized through the quality factor $Q=\omega_0/\gamma_0$. Note that in the following analysis, we assume that the particle's motion is precooled to near the ground state with $n=0.07$.

\fref{fig:feasibility}(e) and (g) show an increase of both the fringe amplitude and wavenumber with the quality factor followed by saturation. A minimal quality factor of about $Q=10^{11}$ is required to obtain an observable fringe amplitude, see \fref{fig:feasibility}(f). Saturation occurs at quality factors larger than $Q=10^{13}$, see \fref{fig:feasibility}(f) and \fref{fig:feasibility}(h), which represents the maximum visibility achievable for the given parameters of the DP protocol. 
%We identify a minimal quality factor of $Q=10^{11}$ that would be required to observe clear quantum features in the motion of a $1$\,\textmu m radius superconducting microparticle.

%%%%%%%%%%%%%%%%%%%%%%%%%%%%%%
%%%%%%%%%%%%%%%%%%%%%%%%%%%%%%%%%%%%%%%%%%%%%%%%%%%%%%%%
\section{Qualitative analysis}\label{sec:analytics}
%%%%%%%%%%%%%%%%%%%%%%%%%%%%%%%%%%%%%%%%%%%%%%%%%%%%%%%%
%%%%%%%%%%%%%%%%%%%%%%%%%%%%%%

To gain insight into the phase-space dynamics beyond the numerical simulations, we employ analytical methods exploiting appropriate approximations to arrive at scaling relations for the fringe amplitude and wavenumber. As a result, we obtain analytic functions of these quantities in dependence of the initial phonon occupation and the quality factor as seen in \fref{fig:feasibility}(d,f,h), but not for \fref{fig:feasibility}(b), see \aref{app:dynamics_ana}. In the following, we will give the idea behind that analytic treatment and state the results, while details of the derivation can be found in \aref{app:dynamics_ana}.

We examine the dynamics of the state generated via \eref{eq:H_quantum_completeDP} and apply to this equation the rotating-wave approximation (RWA) to eliminate off-resonant contributions. We obtain 
\begin{align}
     \hat{\mathcal{H}}_{\mathrm{RWA}}/\hbar = \Omega_0\, \hat{a}^\dagger \hat{a} 
     + \Omega_{\mathrm{sq}} \left(\hat{a}^{\dagger 2} + \hat{a}^2\right) 
     + \Omega_{\mathrm{Kerr}} \left(\hat{a}^\dagger \hat{a}\right)^2,
     \label{eq_qualitative_01}
\end{align}
with a renormalized resonance frequency $\Omega_0 = \omega_0/2 - \Omega_2$, squeezing strength $\Omega_{\mathrm{sq}} = -\left(\omega_0/4 - \Omega_2\right)$, and Kerr nonlinearity $\Omega_{\mathrm{Kerr}} =3\Omega_4/2$. Importantly, the squeezing term is kept from the RWA as $\Omega_{\mathrm{sq}}$ is not generally small compared to the relevant frequency scale set by $\omega_0$, unlike the case considered in, e.g., Ref.~\cite{PRXQuantum.5.030312}.

The dynamics of the particle's motional state $\hat{\rho}$ can be described by the Lindblad master equation 
\begin{align}
    \frac{d}{dt}\hat{\rho}
    = \hat{\mathcal{L}}_\mathrm{sq}[\hat{\rho}] + \hat{\mathcal{L}}_\mathrm{Kerr}[\hat{\rho}],
    \label{Eq002}
\end{align}
which is composed of a dominant quadratic squeezing contribution including the free evolution (also known as generalized squeezing) and a weaker noisy Kerr contribution, which are given by
\begin{align}
    \hat{\mathcal{L}}_{\mathrm{sq}}[\hat{\rho}] &= -i\left[\hat{\mathcal{H}}_{\mathrm{sq}}/\hbar, \hat{\rho} \right], \\
    \hat{\mathcal{L}}_{\mathrm{Kerr}}[\hat{\rho}] &= -i\left[\hat{\mathcal{H}}_{\mathrm{Kerr}}/\hbar, \hat{\rho} \right] \nonumber \\
    &\quad + \hat{\mathcal{D}}[\sqrt{\gamma_0 \bar{n}}\, \hat{a}^\dagger]\hat{\rho} + \hat{\mathcal{D}}[\sqrt{\gamma_0 (\bar{n}+1)}\, \hat{a}]\hat{\rho} ,
    \label{eq_lindblad_split}
\end{align}
with
\begin{align}
    \hat{\mathcal{H}}_{\mathrm{sq}}/\hbar &= \Omega_{\mathrm{sq}} \left(\hat{a}^{\dagger 2} + \hat{a}^2\right)+\Omega_0\, \hat{a}^\dagger \hat{a}, \\
    \hat{\mathcal{H}}_{\mathrm{Kerr}}/\hbar &= \Omega_{\mathrm{Kerr}} \left(\hat{a}^\dagger \hat{a}\right)^2
\end{align}
and the dissipator terms $\hat{\mathcal{D}}[\hat{O}]\hat{\rho}$ are of the standard kind, with their relation to the dissipative Liouvillian term $\mathcal{L}_{\text{n}}$ discussed in \aref{app:lindbladian_equivalence}. The formal solution to \eref{Eq002} is
\begin{align}\label{eq:lindbladsolution}
    \hat{\rho}(t) = e^{(\hat{\mathcal{L}}_{\mathrm{sq}} + \hat{\mathcal{L}}_{\mathrm{Kerr}})t} \hat{\rho}(0).
\end{align}
One could formally solve \eref{eq:lindbladsolution} to obtain the motional state $\hat{\rho}(t)$ of the particle at any time, possibly using the decomposition described in \aref{app:dynamics_ana}. However, the RWA has been invoked such that the solution may deviate from our numerical simulation of the complete phase space dynamics. Furthermore, solving \eref{eq:lindbladsolution} in the Fock basis requires use of a large enough Hilbert space, which is computationally demanding. Therefore, we have chosen to numerically simulate the dynamics in phase space only.

%%%%%%%%%%%%%%%%%%%%%%%%%
\subsection{Fringe amplitude}
%%%%%%%%%%%%%%%%%%%%%%%%%

We start with finding a relation between the fringe amplitude and the dissipation $\gamma_0$. We consider the short time regime $\Omega_\mathrm{Kerr}t\ll 1$, which is applicable to our case, at which fringes already appear in the Wigner function. In that situation, the evolution can be approximately factorized using the Trotter decomposition \cite{Gilles}
\begin{align}
    \hat{\rho}(t) \approx e^{\frac{1}{2}\hat{\mathcal{L}}_\mathrm{sq} t}
    e^{\hat{\mathcal{L}}_\mathrm{Kerr} t}
    e^{\frac{1}{2}\hat{\mathcal{L}}_\mathrm{sq} t}\,\hat{\rho}(0),
    \label{Eq02}
\end{align}
which enables an effective decomposition into simpler sub-dynamical terms of the pure generalized squeezing dynamics given by $e^{\hat{\mathcal{L}}_\mathrm{sq} t}$ and the noisy Kerr evolution given by $e^{\hat{\mathcal{L}}_\mathrm{Kerr} t}$. We note that a more accurate higher-order Trotter decomposition would merely involve a longer sequence of squeezing-Kerr-squeezing terms. 

This decomposition into sub-dynamics provides a deeper analytical insight into the open system dynamics as each of these sub-dynamics have already been studied. The squeezing-induced dynamics $e^{\frac{1}{2}\hat{\mathcal{L}}_\mathrm{sq} t}$ is a generalized squeezing transformation, as has been analyzed in Ref.~\cite{PRXQuantum.6.010201}. In contrast, the noisy Kerr dynamics describes a dissipative anharmonic oscillator and admits an exact Lie-algebraic decomposition as \cite{chaturvedi1991solution}
\begin{align}
    e^{\hat{\mathcal{L}}_\mathrm{Kerr} t}= e^{\hat{F}_c} e^{\hat{F}_+ \hat{K}_+} e^{\hat{F}_3 \hat{K}_3} e^{\hat{F}_- \hat{K}_-},
    \label{eq_qualitative_02}
\end{align}
where the operators $\hat{K}_\bullet$ represent some finer distinct sub-dynamical generators, and the coefficients $\hat{F}_\bullet$ encode their corresponding amplitudes, see \aref{app:dynamics_ana} for their explicit definitions.

Crucially, the term $e^{\hat{F}_3 \hat{K}_3}$ governs the amplitude of the density matrix elements. In particular, it determines the decay of the amplitude of the interference fringes in phase space. An associated effective decay factor $\gamma_{\mathrm{eff}} t$ can be estimated via the truncated operator norm \cite{Winter17,Shirokov_2020,Becker25} in the case of $\gamma_0 t\ll 1$, which holds for the protocols we consider. We obtain
\begin{align}\label{eq:decay_amplitude}
    \gamma_{\mathrm{eff}} t&\equiv \| \hat{F}_3 \hat{K}_3 \|_{z_0} \approx {z_0}^2\gamma_0 t (1 + 2\bar{n})/4,
\end{align}
with the truncation being performed at Fock state number $n_0 \approx {z_0}^2/4$ and $z_0$ denoting the phase-space coordinate at which the fringe amplitude is evaluated. This expression is similar to the one found in Refs.~\cite{PRL_Roda-Llordes2024,RieraCampeny2024analytical}.

We can now use \eref{eq:decay_amplitude} in a fit of an exponential decay $\exp{(-\gamma_{\mathrm{eff}} t)}$ of the fringe amplitude due to coupling to the thermal environment. Note that we fit a prefactor to the exponential and $z_0$, while $\gamma_0$ and $\bar{n}$ are set by the initial conditions and the time $t$ is fixed. The result is shown in \fref{fig:feasibility}(f) in dependence on the quality factor $Q=\omega_0/\gamma_0$. Despite the slight discrepancy in the exact thermal coupling (\aref{app:lindbladian_equivalence}), the functional dependence is correctly captured by the simplified analytical model. Further, we obtain fitted values of ${z_0}=40.3$ and ${z_0}=45.9$ for $\wzini t/2\pi$ of $10$ and $13.28$, respectively. These values are consistent with the state size along the position quadrature as shown in the Wigner function plot in \fref{fig:Wigner_protocols} at similar times. 

Examining \eref{eq:decay_amplitude} indicates that states with fringes appearing further away from the origin, i.e., having larger $z_0$, experience a stronger decay. This implies an experimental trade-off between a faster state generation realized through an increased squeezing rate $\Omega_{\text{sq}}$ and the increased coherence requirement for larger states, and is relevant when comparing the DP and DWP protocols.

%%%%%%%%%%%%%%%%%%%%%%%%%
\subsection{Fringe wavenumber}
%%%%%%%%%%%%%%%%%%%%%%%%%

Next, we study the peak wavenumber $k_{\mathrm{max}}$ of the interference fringes. To this end, we analyze the Wigner function for a density matrix evolving under the Kerr interaction and identify a threshold that determines the dominant oscillations in phase space. The detailed heuristic argument for finding its scaling behavior is given in \aref{app:fringewavevector}. We arrive at
\begin{align}
 n(k_\mathrm{max}) &\approx A e^{\frac{B}{k_\mathrm{max}}} - C k_\mathrm{max},
    \label{eq:scaling_behavior}
\end{align}
where $A$, $B$, and $C$ are fitting parameters characterizing the scaling behavior. We fit \eref{eq:scaling_behavior} to the inverse numerical data, i.e. $n(k_\mathrm{max})$, see \fref{fig:feasibility}(d). We obtain a good correspondence between the analytic scaling and our numerical results.

To arrive at a scaling relation for $k_\mathrm{max}(Q)$, we make the assumption that the dynamics of a thermal state in a closed system [\fref{fig:feasibility}(c,d)] is equivalent to the dynamics of a vacuum state in a hot thermal environment  [\fref{fig:feasibility}(g,h)], i.e., when $k_BT\gg \hbar\omega_0$. In both cases, the Fock state distribution will be broadened by populating higher phonon number states, thereby increasing the mean phonon occupation. Since this redistribution plays an analogous role in both cases, we treat them on equal footing, see \aref{app:fringewavevector}. We arrive at
\begin{align}\label{eq:Qfitt}    
    Q(k_\mathrm{max})
    &\approx -\omega_0 t\left[\ln\left(1 - \frac{A'}{\bar{n}} e^{\frac{B'}{k_\mathrm{max}}} + \frac{C'}{\bar{n}} k_\mathrm{max} \right)\right]^{-1}, 
\end{align}
with $A'$, $B'$, and $C'$ are fitting parameters characterizing the scaling behavior. We fit again to the inverse numerical data of \eref{eq:Qfitt}, i.e., $k_\mathrm{max}(Q)$. The result is shown in \fref{fig:feasibility}(h) and we find a good agreement between the numerical results and the analytic scaling.

%%%%%%%%%%%%%%%%%%%%%%%%%%%%
%%%%%%%%%%%%%%%%%%%%%%%%%%%%
\section{Experimental feasibility}\label{sec:implementation}
%%%%%%%%%%%%%%%%%%%%%%%%%%%%
%%%%%%%%%%%%%%%%%%%%%%%%%%%%

We now discuss the experimental feasibility of realizing the proposed protocols. We place particular focus on the precision and stability required to generate the magnetic potentials, on how to resolve quantum fringes via a measurement of the position of the microparticle, and on how to distinguish quantum from classical dynamics with statistical confidence.

%%%%%%%%%%%%%%%%%%%%%%%%%%%%
\subsection{Magnetic potential landscape}
%%%%%%%%%%%%%%%%%%%%%%%%%%%%

Both the DP and DWP potential require realization of accurate magnetic field ratios. To this end, the current in the field generating coils as well as the coil geometry must be precise. Let us give an estimate on the required accuracy and stability. Assuming a Pb particle of $\Rp=1$\,\textmu m and $\Bho$ such that $\wzini=2\pi\cdot 100$\,Hz with $\Bh=\Bho$, the required field detuning $|d|$ must be smaller than $2.6\cdot 10^{-9}$ to achieve $D_{\text{DWP}}/z_{\text{zpf}} < 100$. This means that we need an accuracy and stability of the magnetic fields on the order of ppb.

Magnetic fields can reach ppb-level stability when sourced by persistent currents in superconducting coils, see, for example, Refs.~\cite{gabrielseSelfshieldingSuperconductingSolenoid1988,Brouwer2022,fanMeasurementElectronMagnetic2023}. Our protocols rely on dynamically changing the magnetic field landscape and, therefore, switching of magnetic fields. We expect maintaining stability during switching to be an additional challenge beyond steady-state stability. Hence, it may be desirable to reduce the number of dynamically switched magnetic fields. A possibility could be, for example, to realize the protocol either by initially having $\Bqo$ turned on, and then going directly to the desired ratio of $B_1$ and $B_3$, or through initially having $\Bho$ turned on, see \tref{tab:parameters}. Then, a protocol starting from $\Bho$ has an advantage over one starting from $\Bqo$ as only one field needs to be dynamically varied. By a suitable implementation of flux pumping \cite{Coombs2019} and superconducting contactors \cite{Ohtsuka1998}, even dynamically controlled coils could operate in persistent current mode, and, thus, reach a ppb-level of stability \cite{Brouwer2022}. Furthermore, the field switching time must be slow enough to avoid the occurrence of displacement currents, but faster than the oscillation period of the particle. The former time scale is set by $\Rp/c$, where $c$ is the speed of light in vacuum. For particle sizes on the order of \textmu m, the switching times must be larger than fs. The particle's motional period is on the order of ms for a trap frequency of $\wzini=2\pi\cdot 100$\,Hz. Hence, a switching time of  \textmu s would be appropriate.

The precise geometry of the coils affects the ability to produce the desired nonlinearity. For the hexapole configuration shown in \fref{fig:coils_and_potentials}(b) with $\Rch=150\Rp$, displacing one of the lateral coils by $\Rp$ toward the trap center produces, within the small-sphere approximation, a double-well potential with a separation $D\simeq\Rch/25$, rather than the quartic single-well potential shown in \fref{fig:coils_and_potentials}(e). This separation $D$ exceeds $z_\text{zpf}$ by several orders of magnitude. Such geometric imperfections can, however, be compensated by fine-tuning currents. In this example, adding a current of approximately $I_\text{3}/77$ to the central coil restores the quartic single-well configuration; see \aref{app:robustness} for details. Wide double-well potentials can also be deliberately engineered by only employing the hexapole configuration and finely tuning the current through the central coil. Such a configuration could for example be used to trap two microparticles \cite{PRL_Roda-Llordes2024}. Small errors in coil orientation or radius produce comparable field distortions. For example, tilting one of the lateral coils by $1^\circ$ increases the ratio $\Omega_2/\Omega_4$, i.e., the harmonic contribution, and shifts the potential minimum by roughly $\Rch/50$. 

We therefore anticipate that the experimental implementation will require a precise tune-up of the coil currents to compensate for geometric inaccuracy, possibly involving the use of additional compensation coils beyond the configuration shown in \fref{fig:coils_and_potentials}. Furthermore, it will be required to use superconducting coils, which are driven by persistent currents to achieve a ppb-level of stability.

%%%%%%%%%%%%%%%%%%%%%%%%%%%%
\subsection{Resolving fringes in position}
%%%%%%%%%%%%%%%%%%%%%%%%%%%%

The fringes in the position probability distribution are typically peaked at wavenumbers $k_{\text{max}}\sim 0.25/z_{\text{zpf}}$, see \fref{fig:transform}(b) and \fref{fig:feasibility}(d, h). This implies that a position resolution smaller than $1/2k_\text{max} = 2z_{\text{zpf}}$ is required to resolve them. 

We consider first a continuous measurement. Such a measurement would be a natural way to perform state initialization at the start of the protocol using measurement-based feedback cooling of the particle's motion. In the case of cooling as well as for resolving the fringes, the measurement will be bounded by the imprecision noise $S_{zz}^{\text{imp}}$-backaction noise $S_{FF}^{\text{ba}}$ product $\sqrt{S_{zz}^{\text{imp}}S_{FF}^{\text{ba}}}=\epsilon\hbar$, with the energy resolution $\epsilon\geq1/2$ and $\epsilon=1/2$ for the standard quantum limit \cite{ClerkNoise2010}. In case of measurement-based cooling, a measurement rate $\Gamma_{\text{meas}}\geq\frac{\bar{n}\gamma_0}{(2n+1)^2-4\epsilon^2}$ \cite{sudhir2016thesis,rossiMeasurementbasedQuantumControl2018,Magrini2021} is required to cool to a phonon occupation $n$, which is bounded from below by $n \geq \epsilon-\frac{1}{2}$. We may consider using the same measurement procedure at the end of the protocol to determine $P(z)$. For resolving fringes with a resolution on the order of $2z_{\text{zpf}}$ for $n\sim 1$, a continuous measurement with a duration of about $\omega_0/(2\pi \Gamma_{\text{meas}})\sim 10^4$ for $Q=10^{11}$, $T=10\,$mK, $\wzini=2\pi\cdot 100\,$Hz, and $\epsilon=1/2$ is required. This is far longer than the state will exist for and, thus, a continuous measurement is unsuitable.

An alternative measurement approach relies on a strong pulsed position measurement in a cavity optomechanics setting \cite{vanner_pulsed, vanner2013, Vanner2015, Muhonen19}. In that case, the motion of the particle is coupled to a cavity, which can be realized through inductive magnetomechanical coupling, see Refs.~ \cite{rodrigues2019, schmidt2020, ZoepflPRL, SchmidtPRApplied}. In the bad cavity limit, i.e., when the cavity decay rate $\kappa\gg\wzini$, a pulsed measurement of the particle probability distribution $P(z)$ leaves an imprint on the phase quadrature $P_L$ of the light field as \cite{vanner_pulsed}
\begin{equation}\label{eq:pulsed_convolution}
    P(P_L)=\frac{1}{\sqrt{2\pi\sigma_{P_L}^2}}\int e^{-(\chi z/ z_{\text{zpf}}-P_L)^2/2\sigma_{P_L}^2}P(z)dz,
\end{equation}
where $\sigma_{P_L}$ is the uncertainty of the input light phase quadrature with  $\sigma_{P_L}=\sqrt{\epsilon}$ for unsqueezed input
and $\chi\propto g_0\sqrt{n_{\text{cav}}}/\kappa$ is the measurement strength with the coupling rate $g_0$ and number of intracavity photons in the pulse $n_{\text{cav}}$. For the measurement to linearly map the position $z$ to $P_L$, it is required that $z/z_{\text{zpf}}\ll \kappa/g_0$ \cite{Muhonen19}. 

\begin{figure}[t!bhp]
    \centering
    \includegraphics[width=\linewidth]{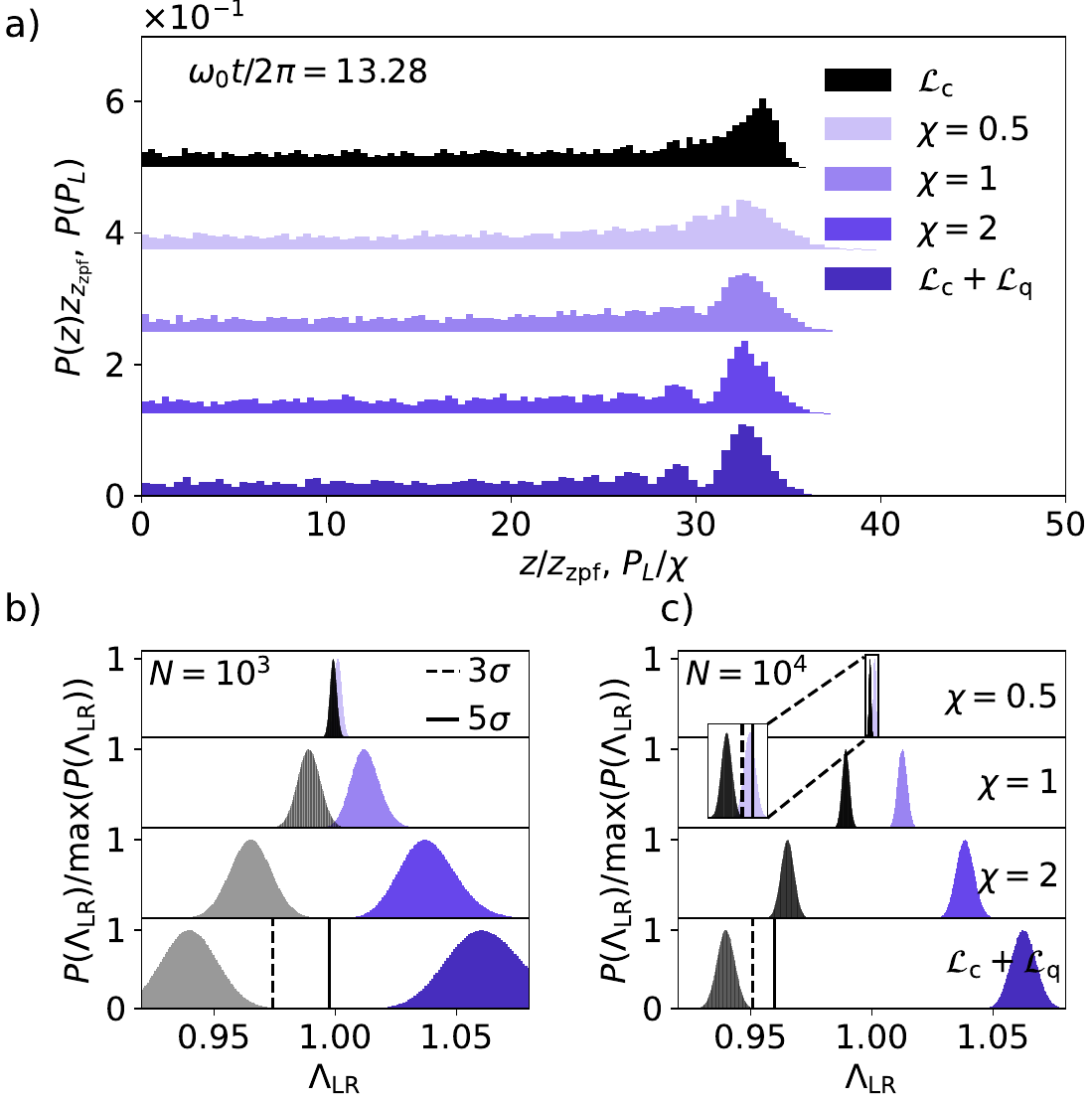}
    \caption{Hypothesis testing for quantum dynamics. (a) Histograms built from $N=10^4$ samples of the position probability distribution for classical dynamics (black), and quantum dynamics when measured through pulsed measurement according to \eref{eq:pulsed_convolution} with $\epsilon=1/2$ (purple). (b,c) Probability distributions for the likelihood ratio under the quantum (purple) and classical (black) hypotheses for the dynamics for (b) $10^3$ and (c) $10^4$ samples. Decision boundaries corresponding to $3\sigma$ (dashed) and $5\sigma$ (solid) are shown. The probability distributions for the likelihood ratio for varying measurement strength $\chi$ are shown stacked in (b,c).}
    \label{fig:sampling}
\end{figure}

\fref{fig:sampling}(a) shows sampled histograms of $P(P_L)$ for various $\chi$. With $\chi\geq1$, the interference fringes can be visually observed in the histograms, while with decreasing $\chi$ they become unresolvable. Current experimental demonstrations of cavity readout of a levitated superconducting microsphere realize $g_0\sqrt{n_{\text{cav}}}/\kappa\sim 10^{-9}$ \cite{SchmidtPRApplied}, and thus, will require substantial improvement in both signal collection and coupling \cite{Paradkar2025}. Optical cavity readout of levitated superconductors could be an alternative \cite{Hansen2026}, but is so far limited by particle heating, preventing long experimental runs, and would require levitation of high-reflectivity objects the shape of which would prevent obtaining the desired trap nonlinearity.

Finally, an option could be to decrease the wavenumber of the generated fringes at the end of the protocol by using an inverted harmonic potential, as, e.g., suggested in Refs.~\cite{romero-isartCoherentInflationLarge2017,QST_Pino2018, Neumeier2024, tomassi26}. Using HOTs up to order three, this can be achieved by suitably tuning $\Bu$ and $\Bh$, in the same manner to what can be done to realize the DWP potential. Then, a coarser measurement resolution could be tolerated, at the cost of a more involved protocol.

%%%%%%%%%%%%%%%%%%%%%%%%%
\subsection{Statistical significance}
%%%%%%%%%%%%%%%%%%%%%%%%%

We give now an estimate of the duration of the experiment necessary to observe nonclassicality with statistical confidence, following  Ref.~\cite{rieracampeny2025certifying}. We consider a simple binary test for distinguishing between a classical and a quantum model for the system dynamics \cite{ralphDynamicalModelSelection2018,rieracampeny2025certifying}. A more complete statistical analysis would also have to take into account uncertainty in the various experimental parameters, which we here consider to be completely known. We define as a test statistic the likelihood ratio between two alternative models $1$ and $0$ as
\begin{equation}
    \Lambda_{\text{LR}}={\left(\prod_{i=1}^N\frac{P_1(d_i}{P_0(d_i)}\right)}^{1/N},
\end{equation}
for the data $D=\{d_i\}_{i=1}^N$ with the individual samples $d_i$ being drawn from a true probability distribution for the experimental outcome, the total number of samples $N$, and the model probability distributions $P_1$ for the alternative hypothesis and $P_0$ for the null hypothesis.

Taking the $n=0$ state shown in \fref{fig:transform}, we compute the likelihood ratio $\Lambda_{\text{LR}}$ with quantum dynamics as the alternative hypothesis $P_1$ and classical dynamics as the null hypothesis $P_0$ for data drawn from the predictions of the respective dynamics. We first consider the limiting cases of sampling from the exact position probability distributions without any resolution limit and compute the likelihood ratio for quantum data ($\mathcal{L}_{\text{q}}+\mathcal{L}_{\text{c}}$) as well as for classical data ($\mathcal{L}_{\text{c}}$). Histograms of these distributions are shown in the bottom part of \fref{fig:sampling}(b) and \fref{fig:sampling}(c) for $N=10^3$ and $10^4$, respectively.

We determine that for $N=10^3$ [\fref{fig:sampling}(b)], the statistical power is near unity for both $3\sigma$ and $5\sigma$. With $\chi=1$, the statistical power at $3\sigma$ ($5\sigma$) decreases to $94\%$ ($36\%$), and decreases further to $5.5\%$ ($0.03\%$) for $\chi=0.5$. An experiment limited in measurement strength would thus be unlikely to confirm quantum dynamics at $5\sigma$ given $10^3$ repetitions. The statistical power increases substantially for $N=10^4$ [\fref{fig:sampling}(c), inset], achieving $3\sigma$ ($5\sigma$) at $90\%$ ($29\%$) for $\chi=0.5$. 

Finally, let us estimate the total time required to perform the experiment. A single run requires
\begin{equation}
    t_{\text{run}}=t_{\text{cool}}+t_{\text{evo}}+t_{\text{meas}},
\end{equation}
with $t_{\text{cool}}\sim 1/\Gamma_{\text{meas}}$, $t_{\text{evo}}\sim 15\cdot 2\pi/\omega_0$, and $t_{\text{meas}}\ll 2\pi/\omega_0$, as we assume an instantaneous pulsed measurement. The cooling time is about 600\,s for $Q=10^{11}$ and $T=10$\,mK and dominates the time scales. Note that with access to pulsed measurement, one could consider pulsed cooling schemes from, e.g., Refs.~\cite{vanner_pulsed,machnesPulsedLaserCooling2012} to decrease the required cooling time drastically. With a pulsed measurement of strength $\chi=1$, the total experimental run time is expected to be seven days when repeating the experiment $N=10^3$ times to acquire statistics that would give a statistical power of $94\%$ at $3\sigma$. During this time the experiment must be sufficiently stable such that the experiment reproduces an identical state up to the measurement resolution. Though this is a challenging experimental task, the time scale would not be unrealistic for an experiment at this level of sophistication.

%%%%%%%%%%%%%%%%%%%%%%%%%%%%%%
%%%%%%%%%%%%%%%%%%%%%%%%%
\section{Conclusions and outlook}\label{sec:theend}
%%%%%%%%%%%%%%%%%%%%%%%%%
%%%%%%%%%%%%%%%%%%%%%%%%%%%%%%

Higher-order magnetic traps provide a powerful and flexible framework for realizing anharmonic trap potentials for magnetically levitated superconducting microparticles. We have shown how to achieve strongly anharmonic potentials such as a double-well or Duffing potential. Crucially, the nonlinearity is already active on the scale of about hundred times the zero-point motion of the particle, which enables generation of non-Gaussian states of the center-of-mass motion of the trapped microparticle with an extent of about 0.1\,nm.

We have numerically simulated the dynamics in phase space of a trapped $1\,$\textmu m radius microparticle exposed to these nonlinear potentials. We identified clear signatures of nonclassicality in the position probability distribution, provided the 
particle's quality factor is at least $10^{11}$ and cooled close to its motional ground state. We derived analytic scaling relations relating the observability of quantum fringes to the quality factor and initial phonon occupation of the particle's motion, which guides the search for relevant parameter regimes.

We have outlined how one can discern classical from quantum dynamics based on the measurement of the position probability distribution of the trapped particle. To this end, we proposed pulsed measurement in a cavity magnetomechanics setting as a viable measurement strategy for position. We showed how the measurement strength of the pulsed measurement would set the number of repetitions of the experiment needed to confirm quantum dynamics at a given significance level and statistical power.

Overall, our work outlines an experimentally challenging path to observe nonclassicality in the center-of-mass motion of a trapped superconducting microparticle with a mass in the picogram regime. Such an experiment would contribute to the pursuit of generating quantum states of objects with increasing size and mass \cite{Arndt1999,oconnellQuantumGroundState2010,kovachyQuantumSuperpositionHalfmetre2015,bildSchrodingerCatStates2023,Pedalino26}. Our proposal leverages the unique opportunities that superconducting magnetic levitation offers: the trap is noiseless as it is driven by superconducting persistent currents, the trap field has no upper mass limit, HOTs provide built-in trap anharmonicity on length scales close to the zero-point motion, and the motion of the microparticle can be coupled to superconducting quantum circuits via flux-based coupling schemes for efficient read-out \cite{ZoepflPRL,rodrigues2019,schmidt2020,SchmidtPRApplied}. Interesting extensions of our proposal would be to consider trapping of nonspherical particles \cite{bort-soldevila2024} in anharmonic magnetic trap potentials and their suitability for the observation of rotational quantum states \cite{sticklerQuantumRotationsNanoparticles2021}, or trapping of multiple particles. The non-Gaussian motional state of the levitated superconducting microparticle could be used in future experiments for sensing tasks \cite{armataQuantumLimitsGravity2017,qvarfortGravimetryNonlinearOptomechanics2018}, for probing unconventional decoherence models \cite{diosiUniversalMasterEquation1987,penroseGravityRoleQuantum1996,collapse_RevModPhys} or testing modifications of quantum mechanics \cite{NimmrichterHornberger, Schrinski2019}, or for exploring the interplay between quantum physics and gravity  \cite{boseSpinEntanglementWitness2017,marlettoGravitationallyInducedEntanglement2017,Aspel2018,Bose25}.

%%%%%%%%%%%%%%%%%%%%%%%%%%%%%%
%%%%%%%%%%%%%%%%%%%%%%%%%%%%%%

\begin{acknowledgments} 
  We acknowledge valuable discussions with Hugo Johansson, Natanael Bort-Soldevila, Nuria Del-Valle, Thomas Agrenius, Andrey Danilov, Nikolai Kiesel and Benjamin Stickler. This work was supported in part by the Horizon Europe 2021-2027 Framework Programme (European Union) through the SuperMeQ project (Grant Agreement number 101080143), the European Research Council under Grant No. 101087847 (ERC Consolidator SuperQLev), and the Knut and Alice Wallenberg (KAW) Foundation through a Wallenberg Academy Scholar (WW). JC-S acknowledges funding from AGAUR-FI Joan Or\'o grants (Grant No. 2025 FI-3 00143) of the Generalitat de Catalunya and the European Social Fund Plus. SQ acknowledges funding by the Wallenberg Initiative on Networks and Quantum Information (WINQ). Nordita is funded in part by NordForsk.
\end{acknowledgments}

%%%%%%%%%%%%%%%%%%%%%%%%%%%%%%
\section*{Author contributions}
%%%%%%%%%%%%%%%%%%%%%%%%%%%%%%

JC-S and FR contributed equally to this work. JC-S, FR, and WW conceptualized the work. JC-S and FR performed the electromagnetic analysis with support from CN. JC-S proposed and investigated the use of magnetic higher-order traps and performed numerical simulations of the trap configurations. FR performed the phase-space analysis, numerically simulated the system dynamics, and carried out the feasibility study. SZ and SQ developed the analytical model of the system dynamics. SQ, AM, CN, and WW supervised the work. JC-S, FR, SZ, and WW wrote the manuscript with input from all authors.

%%%%%%%%%%%%%%%%%%%%%%%%%%%%%%
%%%%%%%%%%%%%%%%%%%%%%%%%%%%%%
%%%%%%%%%%%%%%%%%%%%%%%%%%%%%%
%\clearpage
\appendix
% \counterwithin{figure}{section}
% \renewcommand{\thefigure}{A\arabic{figure}}
% \counterwithin{table}{section}
% \renewcommand{\thetable}{A\Roman{table}}
%\onecolumngrid
\section*{Appendix}

%%%%%%%%%%%%%%%%%%%%%%%%%%%%%%
%%%%%%%%%%%%%%%%%%%%%%%%%%%%%%
\section{Theoretical framework for magnetic HOTs}
%%%%%%%%%%%%%%%%%%%%%%%%%%%%%%
%%%%%%%%%%%%%%%%%%%%%%%%%%%%%%

%%%%%%%%%%%%%%%%%%%%%%%%%%%%%%
\subsection{Magnetic potential energy for an arbitrary magnetic field}\label{app:magnetic_energy}
%%%%%%%%%%%%%%%%%%%%%%%%%%%%%%

The magnetic potential energy---magnetostatic interaction energy---experienced by a magnetic object, with a simply-connected volume $\Vp$ and surface $\Sp$, in the presence of an applied magnetic induction field $\B_\text{a}$ can be calculated as
 \begin{equation}
    \mathcal{U}=-\frac{1}{2}\int_{\Vp}\left(\M\cdot\B_\text{a}\right)\,\diff V,
\end{equation}
where $\M=\M(\H)$ is the magnetization of the material [assumed to be linear to $\H=\H(\B_\text{a})$], and it is induced by the applied field. Here, the back-action of the magnetic object on the currents generating $\B_\text{a}$ is neglected, such that the applied field is assumed to remain unaffected by the presence of the object. The magnetic force on the object will be $\textbf{F}=-\nabla_\xi \mathcal{U}$, with $\xi$ being a small displacement of the object \cite{jackson1999,landau1984}.

If we assume a magnetic object with a linear magnetic susceptibility $\chi_m=-1$ (perfect diamagnetic material), then inside the material,
\begin{equation}
    \M\cdot\B_\text{a}=-\frac{|\B_\text{a}|^2}{\mu_0}-\Hdi\cdot\B_\text{a},
\end{equation}
with $\Hdi=\H_\text{in}-\H_\text{a}$ being the demagnetizing field due to the currents induced on the surface of the object. If there are no free currents, we can define this field as the gradient of a magnetic scalar potential $\Hd\equiv-\nabla\phi_\text{d}$. Then
\begin{align}\label{eq:potential_energy_two_terms}
    \mathcal{U} & =\frac{1}{2\mu_0}\int_{\Vp}|\B_\text{a}|^2\diff V-\frac{1}{2}\int_{\Vp}(\B_\text{a}\cdot\nabla\phii)\diff V\\
        & =\mathcal{U}_1+\mathcal{U}_2,\nonumber
\end{align}
where the first term, $\mathcal{U}_1$, is the magnetic energy of the applied field inside the particle. For the second term, $\mathcal{U}_2$, we will use the boundary condition
\begin{align}\label{eq:Neumann} (\B_\text{a}\cdot\mathbf{e}_{n})|_{\Sp}&=\mu_0(\nabla\phio\cdot\mathbf{e}_{n})|_{\Sp}.
\end{align}
Then, since $\nabla\cdot\B_\text{a}=0$ and carefully applying Gauss's theorem at the surface of the object, it becomes
\begin{align}
    \nonumber\mathcal{U}_2 & =-\frac{1}{2}\oint_{\Sp}\phii\B_\text{a}\cdot\mathbf{e}_{n}\diff S\\
        & =\frac{\mu_0}{2}\oint_{\Sp}\phii\nabla\phio\cdot\mathbf{e}_{n}\diff S,
\end{align}
with $\mathbf{e}_{n}$ the normal vector pointing from inside to outside on the object's surface. The magnetic poles give the difference between $|\nabla\phio-\nabla\phii|\cdot\mathbf{e}_{n}$ at the surface.

Now we assume that the diamagnetic object is a spherical particle (radius $\Rp$) centered at $\textbf{r}=(x,y,z)$. Redefining the coordinates in terms of the center of the particle, $\textbf{r}'$, one can write the magnetic potential energy experienced for this particle under any applied magnetic field as
\begin{eqnarray}\label{eq:potential_energy_two_terms_final}
    \nonumber \mathcal{U}(\textbf{r})  &= &\mathcal{U}_1(\textbf{r})+\mathcal{U}_2(\textbf{r})\\
    \nonumber &=&\frac{1}{2\mu_0}\int_{\Vp}|\B_\text{a}(\textbf{r}+\textbf{r}')|^2\diff V'\\
    &&-\frac{\mu_0}{2}\oint_{\Sp}\left(\phii\nabla\phio\cdot\mathbf{e}_{r'}\right)\diff S'.
\end{eqnarray}
The scalar potentials $\phii$ and $\phio$ satisfy Laplace's equation in spherical coordinates. Then, imposing a finite value for them at $r'=0$ and fixing the origin of the potential at $r'\rightarrow \infty$ (since they are the potentials associated with the demagnetizing field, not with the total field),
\begin{equation}
\begin{split}
    &\phii(\textbf{r}')=\sum_{n=0}^{\infty}r'^{n}\sum_{m=-n}^{n}b_{nm}Y_{n}^{m}(\theta', \varphi'),\\
    &\phio(\textbf{r}')=\sum_{n=0}^{\infty}r'^{-(n+1)}\sum_{m=-n}^{n}a_{nm}Y_{n}^{m}(\theta', \varphi'),
\end{split}
\end{equation}
where $Y_n^m(\theta',\varphi')$ are the orthonormalized Laplace spherical harmonics.
The continuity of the potential at the surface of the sphere implies that
\begin{equation}
    b_{nm}=\Rp^{-(2n+1)}a_{nm}.
\end{equation}
Then, the second term of \eref{eq:potential_energy_two_terms_final} can be developed by taking advantage that our potentials are real (i.e. $\phii=\phii^*$), and using that at the surface $\text{d}S'=\Rp^2\sin\theta'\text{d}\theta'\text{d}\varphi'=\Rp^2\text{d}\Omega'$,
\begin{widetext}
\begin{align}
    \mathcal{U}_2 & =-\frac{\mu_0}{2}\sum_{n=0}^{\infty}\sum_{m=-n}^{n}\sum_{n'=0}^{\infty}\sum_{m'=-n'}^{n'}\bigg[ a_{nm}^*a_{n'm'}\left(-n'-1\right)\Rp^{-(n+n'+1)}\oint_{\Sp} {{Y^{m}_n}}^*(\theta',\varphi')Y_{n'}^{m'}(\theta',\varphi')\diff\Omega'\bigg].
\end{align}
\end{widetext}
Finally, by the orthogonality of the spherical harmonics, the potential magnetic energy for the spherical particle becomes \eref{eq:full_pot}.

Note that the coefficients $a_{nm}$ can be easily obtained by using the Neumann boundary condition, as \eref{eq:Neumann},
\begin{equation}\label{eq:Neumann_condition}
    \partial_{r'}\phio\big|_{r'=\Rp}=\mu_0^{-1}(\B_\text{a}\cdot\mathbf{e}_{r'})\big|_{r'=\Rp}.
\end{equation}
Therefore, for a general magnetic induction field the coefficients $a_{nm}(\textbf{r})$ can be obtained as given in \eref{eq:coefficients}.

Finally, and therein lies the strength of this method, one only needs to calculate the coefficients $a_{nm}(\textbf{r})$ by \eref{eq:coefficients} to get the magnetic potential energy of the spherical particle from \eref{eq:full_pot}.

%%%%%%%%%%%%%%%%%%%%%%%%%%%%%%
\subsection{Axisymmetric multipole fields and projection coefficients}\label{app:fields_and_coeff}
%%%%%%%%%%%%%%%%%%%%%%%%%%%%%%

We state here the coefficients of \eref{eq:coefficients} for several applied fields. Centering the spherical particle at $\textbf{r}=(x,y,z)$, we will use the coordinates defined in \sref{sec:HOTS} of the main text. We denote the field $\B_w$ with a numerical subscript $w$, which indicates the order of the field.

The simplest case is that of an external uniform field (order 1), which, notably, does not qualify as a trapping field,
\begin{equation}\label{eq:uniform_field}
    \B_{\text{1}}=b_\text{1}\mathbf{e}_{z}.
\end{equation}
This field produces a response from the spherical diamagnet of
\begin{equation}\label{eq:coeff_uniform}
    a_{10}=-\frac{b_\text{1}}{\mu_0}\sqrt{\frac{\pi}{3}}\Rp^3.
\end{equation}

At order 2, we encounter the external quadrupolar field,
\begin{equation}\label{eq:quadrupole_field}
    \B_{\text{2}}=\frac{b_\text{2}}{2}\left[-(x+x')\mathbf{e}_{x}-(y+y')\mathbf{e}_{y}+2(z+z')\mathbf{e}_{z}\right],
\end{equation}
which produces the response \cite{Hofer2019}
\begin{subequations}\label{eq:coeff_quadrupole}
    \begin{equation}
        a_{1-1}=\frac{b_\text{2}}{\mu_0}\sqrt{\frac{\pi}{24}}\Rp^3(x+iy)\, ,
    \end{equation}
    \begin{equation}
        a_{10}=-\frac{b_\text{2}}{\mu_0}\sqrt{\frac{\pi}{3}}\Rp^3z\, ,
    \end{equation}
    \begin{equation}
        a_{11}=-\frac{b_\text{2}}{\mu_0}\sqrt{\frac{\pi}{24}}\Rp^3(x-iy)\, ,
    \end{equation}
    \begin{equation}
        a_{20}=-\frac{b_\text{2}}{\mu_0}\sqrt{\frac{4\pi}{45}}\Rp^5.
    \end{equation}
\end{subequations}

An external hexapolar field (order 3),
\begin{eqnarray} \label{eq:hexapole_field}
    \nonumber \B_{\text{3}} & =&b_\text{3}\Big[-(z+z')(x+x')\mathbf{e}_{x}-(z+z')(y+y')\mathbf{e}_{y}\\
    &&+\left((z+z')^2-\frac{(x+x')^2+(y+y')^2}{2}\right)\mathbf{e}_{z}\Big],
\end{eqnarray}
generates the response coefficients
\begin{subequations}\label{eq:coeff_hexapole}
    \begin{equation}
        a_{1-1}=\frac{b_\text{3}}{\mu_0}\sqrt{\frac{\pi}{6}}\Rp^3(x+iy)z\, ,
    \end{equation}
    \begin{equation}
        a_{10}=\frac{b_\text{3}}{\mu_0}\sqrt{\frac{\pi}{12}}\Rp^3\left(x^2+y^2-2z^2\right)\, ,
    \end{equation}
    \begin{equation}
        a_{11}=-\frac{b_\text{3}}{\mu_0}\sqrt{\frac{\pi}{6}}\Rp^3(x-iy)z\, ,
    \end{equation}
    \begin{equation}
        a_{2-1}=\frac{b_\text{3}}{\mu_0}\sqrt{\frac{8\pi}{135}}\Rp^5(x+iy)\, ,
    \end{equation}
    \begin{equation}
        a_{20}=-\frac{b_\text{3}}{\mu_0}\sqrt{\frac{16\pi}{45}}\Rp^5z\, ,
    \end{equation}
    \begin{equation}
        a_{21}=-\frac{b_\text{3}}{\mu_0}\sqrt{\frac{8\pi}{135}}\Rp^5(x-iy)\, ,
    \end{equation}
    \begin{equation}
        a_{30}=-\frac{b_\text{3}}{\mu_0}\sqrt{\frac{\pi}{28}}\Rp^7\, .
    \end{equation}
\end{subequations}

Together with the potential energy \eref{eq:full_pot}, these coefficients make it possible to analyze the potential generated for a combination of multipolar fields up to order 3 with fully tunable relative field magnitudes. Note that the units of $\bu$, $b_\text{2}$, $\bh$ are T, T/m, T/m$^2$, respectively. We also use the field magnitudes as $\Bu=\bu$, $\Bq=\bq\Rp$, and $\Bh=\bh\Rp^2$ such that the units of the capital letters are all in T.

%%%%%%%%%%%%%%%%%%%%%%%%%%%%%%
\subsection{Magnetic HOTs beyond the hexapoles}\label{app:beyond_hexa}
%%%%%%%%%%%%%%%%%%%%%%%%%%%%%%

We explore some HOTs of order higher than the hexapole. These traps can be useful for creating a wide variety of potential landscapes beyond the double well or the Duffing potential. In \tref{tab:magnetic_HOT_fields}, derived from Table I in Ref. \cite{PRA_Bergeman1987}, we show the applied HOT cylindrical fields for trapping particles in cylindrical coordinates $(\tilde{\rho},\tilde{\varphi},\tilde{z})$.

\begin{table*}[t!hbp]
\centering
\caption{Magnetic HOT applied fields in cylindrical coordinates, up to the tetrakaidecapolar field. Note that the units of $b_w$ are $\text{T}/\text{m}^{w-1}$, being $w$ the order of the field.}
\label{tab:magnetic_HOT_fields}
\begin{tabular}{c|c|c}
\hline \hline
Order & Magnetic HOT & Trapping field  \\\hline
3 & Hexapole & $\B_{\text{3}}=b_\text{3}\left[-\tilde{z}\tilde{\rho}\mathbf{e}_{\tilde{\rho}
}+\left(\tilde{z}^2-\tilde{\rho}^2/2\right)\mathbf{e}_{z}\right]$\\
4 & Octupole & $\B_{\text{4}}=\bo\left[\left(3\tilde{\rho}^3/8-3\tilde{z}^2\tilde{\rho}/2\right)\mathbf{e}_{\tilde{\rho}
}+\left(\tilde{z}^3-3\tilde{z}\tilde{\rho}^2/2\right)\mathbf{e}_{z}\right]$\\
5 & Decapole & $\B_{\text{5}}=\bd\left[\left(3\tilde{\rho}^3\tilde{z}/2-2\tilde{\rho} \tilde{z}^3\right)\mathbf{e}_{\tilde{\rho}
}+\left(\tilde{z}^4-3\tilde{z}^2\tilde{\rho}^2+3\tilde{\rho}^4/8\right)\mathbf{e}_{z}\right]$\\
6 & Dodecapole & $\B_{\text{6}}=\bdo\left[\left(15\tilde{\rho}^3 \tilde{z}^2/4-5\tilde{\rho} \tilde{z}^4/2-5\tilde{\rho}^5 /16\right)\mathbf{e}_{\tilde{\rho}
}+\left(\tilde{z}^5-5\tilde{z}^3\tilde{\rho}^2+15\tilde{\rho}^4 \tilde{z}/8\right)\mathbf{e}_{z}\right]$\\
7 & Tetrakaidecapole & \small$\B_{\text{7}}=b_\text{7}\left[\left(15\tilde{z}^3\tilde{\rho}^3/2-15\tilde{z}\tilde{\rho}^5/8-3\tilde{z}^5\tilde{\rho}\right)\mathbf{e}_{\tilde{\rho}
}+\left(\tilde{z}^6-15\tilde{z}^4\tilde{\rho}^2/2+45\tilde{z}^2\tilde{\rho}^4/8-5\tilde{\rho}^6/16\right)\mathbf{e}_{z}\right]$\\
\hline \hline
\end{tabular}
\end{table*}

We can now calculate the projection coefficients from \eref{eq:coefficients}. The response coefficients generated by the superconducting sphere for the octupolar field $\B_\text{4}$ and the decapolar field $\B_\text{5}$, from \tref{tab:magnetic_HOT_fields}, are shown in \tref{tab:magnetic_HOT_coeff}. Note that for a cylindrical multipole of order $w$, the only non-zero $a_{nm}$ coefficients are the ones with $1\leq n\leq w$ and $-k\leq m_{(n=w-k)}\leq k$ when they exist ($n,k\in \mathbb{N}_0$, so, when $n\geq k$). The proper combination of these HOT fields allows us to tune the potential landscape for the levitated particle, with each field contributing two additional degrees to the polynomial in terms of the $z$ motion compared to its previous-order field.

\begin{table*}[t!hbp]
\centering
\caption{Projection coefficients for the octupolar and the decapolar field. All coefficients that are not listed in this table are zero for both cases.
}
\label{tab:magnetic_HOT_coeff}
\begin{tabular}{c|cc}
\hline \hline
$a_{nm}(\textbf{r})$ & Octupole & Decapole  \\\hline
$a_{1-1}$ & $-\frac{\bo}{\mu_0}\sqrt{\frac{3\pi}{128}}\Rp^3(x+iy)(x^2+y^2-4z^2)$ & $-\frac{\bd}{\mu_0}\sqrt{\frac{\pi}{24}}\Rp^3(x+iy)z(3x^2+3y^2-4z^2)$ \\
$a_{10}$ & $\frac{\bo}{\mu_0}\sqrt{\frac{\pi}{12}}\Rp^3(3x^2+3y^2-2z^2)z$ & $-\frac{\bd}{\mu_0}\sqrt{\frac{\pi}{192}}\Rp^3\left(3(x^2+y^2)^2-24(x^2+y^2)z^2+8z^4\right)$ \\
$a_{11}$ & $\frac{\bo}{\mu_0}\sqrt{\frac{3\pi}{128}}\Rp^3(x-iy)(x^2+y^2-4z^2)$ & $\frac{\bd}{\mu_0}\sqrt{\frac{\pi}{24}}\Rp^3(x-iy)z(3x^2+3y^2-4z^2)$ \\
$a_{2-2}$ & $-\frac{\bo}{\mu_0}\sqrt{\frac{\pi}{120}}\Rp^5(x+iy)^2$ & $-\frac{\bd}{\mu_0}\sqrt{\frac{2\pi}{15}}\Rp^5(x+iy)^2z$ \\
$a_{2-1}$ & $\frac{\bo}{\mu_0}\sqrt{\frac{8\pi}{15}}\Rp^5(x+iy)z$ & $-\frac{\bd}{\mu_0}\sqrt{\frac{2\pi}{15}}\Rp^5(x+iy)(x^2+y^2-4z^2)$ \\
$a_{20}$ & $\frac{\bo}{\mu_0}\sqrt{\frac{\pi}{5}}\Rp^5(x^2+y^2-2z^2)$ & $-\frac{\bd}{\mu_0}\sqrt{\frac{16\pi}{45}}\Rp^5\left(-3(x^2+y^2)+2z^2\right)z$ \\
$a_{21}$ & $-\frac{\bo}{\mu_0}\sqrt{\frac{8\pi}{15}}\Rp^5(x-iy)z$ & $\frac{\bd}{\mu_0}\sqrt{\frac{2\pi}{15}}\Rp^5(x-iy)(x^2+y^2-4z^2)$ \\
$a_{22}$ & $-\frac{\bo}{\mu_0}\sqrt{\frac{\pi}{120}}\Rp^5(x-iy)^2$ & $-\frac{\bd}{\mu_0}\sqrt{\frac{2\pi}{15}}\Rp^5(x-iy)^2z$ \\
$a_{3-2}$ & $0$ & $-\frac{\bd}{\mu_0}\sqrt{\frac{27\pi}{1120}}\Rp^7(x+iy)^2$ \\
$a_{3-1}$ & $\frac{\bo}{\mu_0}\sqrt{\frac{27\pi}{448}}\Rp^7(x+iy)$ & $\frac{\bd}{\mu_0}\sqrt{\frac{27\pi}{28}}\Rp^7(x+iy)z$ \\
$a_{30}$ & $-\frac{\bo}{\mu_0}\sqrt{\frac{9\pi}{28}}\Rp^7z$ & $\frac{\bd}{\mu_0}\sqrt{\frac{9\pi}{28}}\Rp^7(x^2+y^2-2z^2)$ \\
$a_{31}$ & $-\frac{\bo}{\mu_0}\sqrt{\frac{27\pi}{448}}\Rp^7(x-iy)$ & $-\frac{\bd}{\mu_0}\sqrt{\frac{27\pi}{28}}\Rp^7(x-iy)z$ \\
$a_{32}$ & $0$ & $-\frac{\bd}{\mu_0}\sqrt{\frac{27\pi}{1120}}\Rp^7(x-iy)^2$ \\
$a_{4-1}$ & $0$ & $\frac{\bd}{\mu_0}\sqrt{\frac{64\pi}{1125}}\Rp^9(x+iy)$ \\
$a_{40}$ & $-\frac{\bo}{\mu_0}\sqrt{\frac{4\pi}{225}}\Rp^9$ & $-\frac{\bd}{\mu_0}\sqrt{\frac{64\pi}{225}}\Rp^9 z$ \\
$a_{41}$ & $0$ & $-\frac{\bd}{\mu_0}\sqrt{\frac{64\pi}{1125}}\Rp^9(x-iy)$ \\
$a_{50}$ & $0$ & $-\frac{\bd}{\mu_0}\sqrt{\frac{\pi}{99}}\Rp^{11}$ \\
\hline \hline
\end{tabular}
\end{table*}

%%%%%%%%%%%%%%%%%%%%%%%%%%%%%%
%%%%%%%%%%%%%%%%%%%%%%%%%%%%%%
\section{HOT coils and Hamiltonians}
%%%%%%%%%%%%%%%%%%%%%%%%%%%%%%
%%%%%%%%%%%%%%%%%%%%%%%%%%%%%%

%%%%%%%%%%%%%%%%%%%%%%%%%%%%%%
\subsection{Coil configurations and field strengths}
%%%%%%%%%%%%%%%%%%%%%%%%%%%%%%
For the generation of a quadrupole field ($w=2$), which produces a harmonic trap suitable for ground-state cooling, an anti-Helmholtz coil (AHC) geometry can be used, among other approaches \cite{PRL_Romero-Isart2012,Latorre2020, Cunill-Subiranas2026}. Such an AHC is realized through two coaxial coils with radius $R_\text{c2}$ and $N_\text{2}$ turns, which are separated by $R_\text{c2}$ and carry the same current $I_\text{2}$ in opposite directions, see \fref{fig:coils_and_potentials}(a). Then, the field strength from \eref{eq:quadrupole_field} is given by
\begin{equation}\label{eq:B2}
    \bq=\frac{48\mu_0N_\text{2}I_\text{2}}{25\sqrt{5}R_\text{c2}^2}\implies \Bq=\frac{48\mu_0N_\text{2}I_\text{2}}{25\sqrt{5}}\frac{\Rp}{R_\text{c2}^2}\, .
\end{equation}

By changing the direction of the current threading one of the two coils, i.e., a Helmholtz coil geometry [see \fref{fig:coils_and_potentials}(f,i)], we can directly pass from a quadrupole field to a uniform field, \eref{eq:uniform_field}, with
\begin{equation}\label{eq:Helmholtz_coils_strength}
    \bu=\Bu=\left(\frac{4}{5}\right)^{3/2}\frac{\mu_0 N_\text{1}I_\text{1}}{R_\text{c1}},
\end{equation}
where $N_\text{1}$, $I_\text{1}$, and $R_\text{c1}$ represent the number of turns, current, and radius of the Helmholtz coil. These parameters could take the same values as for the quadrupole if it is the same coil.

For the hexapole field, we propose a spherical coil configuration consisting of three coaxial coils with the same number of turns, $\Nh$. The central coil has radius $\Rch$ and carries a current of $\Ih$. The other two coils with a radius of $\Rch/\sqrt{2}$ are coaxial with the central one, placed at a distance $\pm\Rch/\sqrt{2}$ from it, and carry a current of $\Ih$ in the opposite direction to that of the central coil \cite{PRA_Bergeman1987,PRA_Weinstein1995}. The resulting hexapolar-field strength is then related to the coil current as \cite{PRA_Bergeman1987},
\begin{equation}\label{eq:Bh}
    \bh=\frac{15\mu_0 \Nh\Ih}{8 \Rch^3}\implies \Bh=\frac{15\mu_0 \Nh\Ih}{8}\frac{\Rp^2}{\Rch^3}.
\end{equation}

%%%%%%%%%%%%%%%%%%%%%%%%%%%%%%
\subsection{Realizable Hamiltonians}\label{app:hamiltonians}
%%%%%%%%%%%%%%%%%%%%%%%%%%%%%%

We show some examples of interesting Hamiltonians that one could get from \eref{eq:complete_hamiltonian}. To identify the Hamiltonian resulting from a combination of multipolar fields, we use a vector subscript in which each component denotes the order of the multipoles used. Thus, the Hamiltonian $\mathcal{H}_{(i,j,k)}$ is obtained by combining three cylindrical multipoles of orders $w = i$, $j$, and $k$. Therefore, from \eref{eq:complete_hamiltonian}, quadrupolar traps ($w=2$ only) are governed by the harmonic Hamiltonian,
\begin{equation}
    \mathcal{H}_\text{2}= \frac{|\textbf{p}|^2}{2m}+\frac{\pi \bq^2\Rp^3}{12\mu_0}\left[3\left(x^2+y^2+4z^2\right)+4\Rp^2\right]\, ,
\end{equation}
derived from \eref{eq:quadrupole_field}, \eref{eq:full_pot}, and \eref{eq:coefficients}. Under the small-sphere approximation, $\Rp\ll r_\text{rms}$, the lowest HOT, a magnetostatic spherical hexapole, \eref{eq:hexapole_field}, gives a clearly anharmonic Hamiltonian,
\begin{equation}\label{eq:H_pointparticle}
    \mathcal{H}_{\text{3,small}}=\frac{|\textbf{p}|^2}{2m}+\frac{\pi b_\text{3}^2 R_\text{p}^3}{\mu_0}\left(z^4+\frac{x^4+2x^2y^2+y^4}{4}\right)\, .
\end{equation}
However, for not-so-small particles, the potential landscape is modified accordingly, following \eref{eq:full_pot},
\begin{eqnarray}\label{eq:H_hexapole}
    \nonumber \mathcal{H}_\text{3} & =\frac{\textbf{|p|}^2}{2m}+\frac{\pi b_\text{3}^2}{36\mu_0}\Big[&9\Rp^3(x^4+2x^2y^2+y^4+4z^4)\\
    &&+16\Rp^5(x^2+y^2+3z^2)+6\Rp^7\Big].
\end{eqnarray}
Note that, depending on the motional amplitude, the previously described pure anharmonic potential behaves like a harmonic potential if the amplitude is sufficiently small. This introduces an additional degree of freedom: for \textit{very} large particles ($\Rp\gg z_\text{rms}$), the pure hexapole resembles a harmonic trap, whereas for small particles ($\Rp\ll z_\text{rms}$) the anharmonic term dominates, approaching the small-sphere limit, $\lim_{(\Rp/z_\text{rms})\to0}\mathcal{H}_\text{3}=\mathcal{H}_{\text{3,small}}$.

Hexapolar traps, compared to quadrupolar traps, allow the possibility to create double-well potentials when combined with uniform fields, as previously discussed and shown in \fref{fig:coils_and_potentials}(f). In such cases, the Hamiltonian becomes
\begin{equation}\label{eq:H_combo}
    \mathcal{H}_\text{(1,3)}=\mathcal{H}_\text{3}+\frac{\pi \Rp^3}{\mu_0}\left[\bu^2-\bu\bh \left(x^2+y^2-2z^2\right)\right]\, ,
\end{equation}
so,
\begin{eqnarray}\label{eq:H_double-well}
    \nonumber \mathcal{H}_{z,\text{(1,3)}} & = &\frac{p_z^2}{2m}+\frac{\pi\bh^2\Rp^3}{\mu_0}z^4\\
    \nonumber &&+\frac{2\pi\bh\Rp^3(3\bu+2\bh\Rp^2)}{3\mu_0}z^2\\
    &&+\frac{\pi\Rp^3(6\bu^2+\bh^2\Rp^4)}{6\mu_0}\, .
\end{eqnarray}

Finally, regarding combinations of hexapoles and uniform fields with quadrupole fields, as shown in \eref{eq:complete_hamiltonian}, as long as the axis of the multipoles remains aligned, we may focus on the $z$-motion, which decouples from the $x$ and $y$ motion, and write the Hamiltonian solely restricted to $z$ as
\begin{eqnarray}\label{eq:H_z,uqh}
        \nonumber \mathcal{H}_{z,\text{(1,2,3)}}&=\frac{p_z^2}{2m}+\frac{\pi\Rp^3}{6\mu_0}\bigg[&\left(6\bh^2\right)z^4+\left(12\bh\bq\right)z^3\\
        \nonumber &&+\left(12\bh\bu+6\bq^2+8\Rp^2\bh^2\right)z^2\\
        \nonumber &&+\left(12\bq\bu+8\Rp^2\bh\bq\right)z\\
        &&+6\bu^2+2\Rp^2\bq^2+\Rp^4\bh^2\bigg].
\end{eqnarray}

%%%%%%%%%%%%%%%%%%%%%%%%%%%%%%
\subsection{Overdetermined system of rates $\Omega_j$}\label{app:overdetermined_system}
%%%%%%%%%%%%%%%%%%%%%%%%%%%%%%

The system of equations \eref{eq:rates_short} is overdetermined. This means that not every combination of the four rates is achievable by adjusting the three fields. For example, in \eref{eq:rates_short}, when leaving $\Omega_4$ completely undetermined, the lower-order rates must satisfy
\begin{equation}
    \frac{\Omega_2\Omega_3\pm\sqrt{\Omega_3^2(\Omega^2_2-2\Omega_1\Omega_3)}}{\Omega_1}\geq 0\, \text{ and }\, \Omega_2^2\geq2\Omega_1\Omega_3\, .
\end{equation}
In general, adding the next multipole term $B_w$ to a field that initially includes components up to order $B_{w-1}$  introduces one additional degree of freedom (an extra variable for tuning the rates), while increasing the total number of rates (different from zero) to $2w-2$.

%%%%%%%%%%%%%%%%%%%%%%%%%%%%%%
\subsection{Rates $\Omega_j$ for protocols when cooling in $\Bho$}\label{app:omegas_p2}
%%%%%%%%%%%%%%%%%%%%%%%%%%%%%%

In protocols DP and DWP, when we keep $\Bh=\Bh^\text{ini}$, we calculate the coefficients as 
\begin{align}
    \nonumber \Omega_2 &=\frac{2\pi\Bho\Rp(3\Bu+2\Bho)z_{\text{zpf}}^2}{3\mu_0\hbar}=\sqrt{\frac{1}{2\mu_0\varrho}}\frac{3\Bu+2\Bho}{4\Rp},\\
    \Omega_4 &=\frac{\pi(\Bho)^2z_{\text{zpf}}^4}{\hbar\mu_0\Rp}=\frac{9\hbar}{128\pi\varrho\Rp^5},
\end{align}
and, therefore,
\begin{align} \label{eq:Omegas_from_hexapole}
    \nonumber \Omega_2/\omega_0 &=\frac{3}{8}\frac{\Bu}{\Bho}+\frac{1}{4},\\
    \Omega_4/\omega_0 &=\frac{9\hbar}{128\pi \Rp^4 \Bho}\sqrt{\frac{\mu_0}{2\varrho}}.
\end{align}

%%%%%%%%%%%%%%%%%%%%%%%%%%%%%%
\subsection{Displacement of particle in DWP}\label{sec:displaceDP}
%%%%%%%%%%%%%%%%%%%%%%%%%%%%%%

To achieve an initial displacement of the microparticle in DWP-$\Bqo$, the $xy$-plane of a quadrupole coil system could be intentionally translated by $z_\text{s}$ relative to the coil system generating the double-well. Since achieving geometrical precision at $z_{\text{zpf}}$-level would be difficult, as an alternative uniform fields could be superimposed on the original quadrupole field to spatially shift the position of the potential minimum. This inherent ability to position the particle without dynamic manipulation is an advantage of the D-DWP protocol. Another option would be to initiate the protocol using only two consecutive coils out of the three that generate the hexapole field. In this way, a distorted quadrupolar field would be produced, resulting in a displaced potential minimum. Then, by activating the third coil and the Helmholtz pair responsible for generating the double-well, the particle would initialize on only one side of the double-well, but not at a well-determined location. 

%%%%%%%%%%%%%%%%%%%%%%%%%%%%%%
\subsection{Simulations on required geometric alignment accuracy} \label{app:robustness}
%%%%%%%%%%%%%%%%%%%%%%%%%%%%%%

We analyze the robustness of the trap potential against geometric imperfections of the coil configuration. To this end, we use numerical simulations and quantify the extent to which such imperfections can be compensated by current tuning.

We model the magnetic field generated by the coil geometry shown in \fref{fig:coils_and_potentials}(b), solving Amp\`ere's law for the whole system using finite-element simulations. Each coil is treated as a circular current loop. The effective trap potential along the axial direction $z$ is extracted from the applied field magnitude following the small-sphere approximation, $\mathcal{U}(z)=\frac{\pi\Rp^3}{\mu_0}|\B_\text{a}(z)|^2$.

We start from the ideal, symmetric hexapole configuration with current ratios chosen as described in \fref{fig:coils_and_potentials}. We assume $\Rch=$ \SI{150}{\micro\metre}, $I_\text{3}=1$~A, and a particle with $\Rp=$ \SI{1}{\micro\metre}. Note, however, that the absolute values of these parameters are not essential due to the linearity of the system. Only their relative ratios matter. Small deviations are then introduced in a controlled manner, including lateral displacements or rotations about the coil symmetry axis. For each perturbed configuration, the axial potential is numerically calculated around the trap center. 

\begin{figure}[t!hbp]
    \centering
    \includegraphics[width=\linewidth]{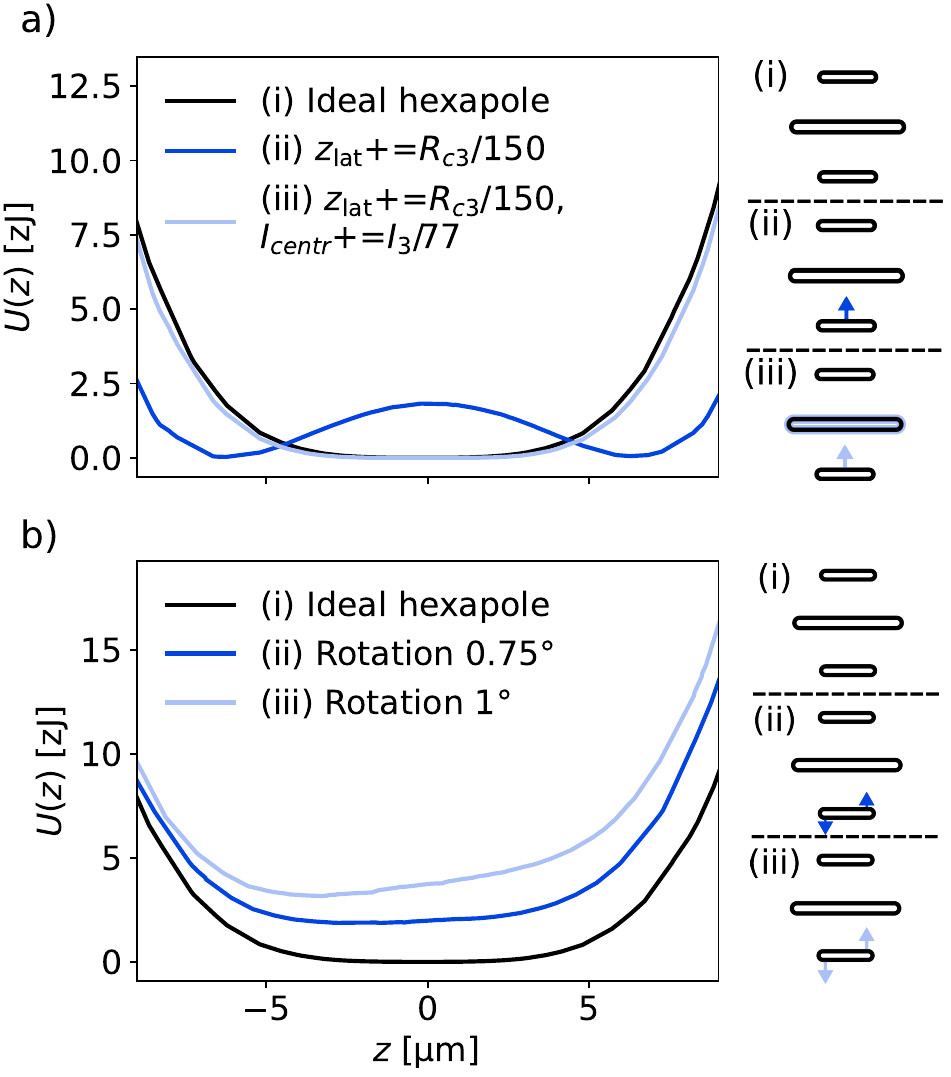}
    \caption{Trap potential along the $z$-axis, under the small-sphere approximation in the presence of geometric imperfections of the hexapolar coils. (a) Potential for a configuration in which one lateral hexapole coil is displaced by \SI{1}{\micro\metre} toward the trap center, resulting in a symmetric double-well potential (blue). A compensated potential can be obtained by increasing the current through the central coil by \SI{13}{\milli\ampere} (light blue), which cancels the induced quadratic contribution. The ideal case is also shown (black). (b) Potential for configurations in which one lateral hexapole coil is rotated about its symmetry axis by $0.75^\circ$ (blue) or $1^\circ$ (light blue), illustrating the sensitivity of the potential shape to small orientation errors.}
    \label{fig:appendix-coils}
\end{figure}

\fref{fig:appendix-coils}(a) shows the trap potential when a vertical displacement of one of the lateral coils by \SI{1}{\micro\metre} toward the trap center is assumed. Then, a symmetric double-well potential is formed. By adjusting the current through the central coil by an amount of \SI{13}{\milli\ampere}, the quadratic contribution induced by the geometric imperfection can be canceled. \fref{fig:appendix-coils}(b) shows the trap potential when one of the lateral hexapole coils has its symmetry axis tilted from the $z$-axis by $0.75^\circ$ or $1.00^\circ$.

%%%%%%%%%%%%%%%%%%%%%%%%%%%%%%
%%%%%%%%%%%%%%%%%%%%%%%%%%%%%%
\section{Phase space dynamics}\label{app:dynamics}
%%%%%%%%%%%%%%%%%%%%%%%%%%%%%%
%%%%%%%%%%%%%%%%%%%%%%%%%%%%%%

We analyze the phase space dynamics using the Wigner function $W(z,p,t)$, which for the classical case is equivalent to the joint probability distribution of finding the particle at position $z$ and momentum $p$ at time $t$ \cite{gardiner-zoller2004}. Under classical dynamics, the Wigner function is a positive-valued function and we can identify the Wigner function with the probability distribution function that is solved for in classical stochastic dynamics \cite{risken1985}, while in the quantum case it can take negative values and $W$ becomes a quasi-probability distribution. However, the marginalized probability distributions for the conjugate variables
\begin{align}
    P(z,t) & =\int W(z,p,t)dp,\\
    P(p,t) & =\int W(z,p,t)dz
\end{align}
are true probability distribution functions in either case. For the quantum case, the Wigner quasiprobability function can also be obtained from the density matrix through
\begin{equation}
    W(z,p,t)=\frac{1}{2\pi\hbar}\int \left<{z+\frac{y}{2}}\right|\hat{\rho}(t)\left|{z-\frac{y}{2}}\right>e^{-ipy/\hbar}dy,
\end{equation}
or the density matrix from a Wigner function through
\begin{equation}
    \hat{\rho}(t)=\int\left[\int\left|{z+\frac{y}{2}}\right>W(z,p,t)e^{ipy/\hbar}\left<{z-\frac{y}{2}}\right|dzdp\right]dy
\end{equation}
\cite{curtright_fairlie_zachos_2013}.

%%%%%%%%%%%%%%%%%%%%%%%%%%%%%%
\subsection{Solving the classical dynamics using Liouvilles theorem}\label{app:liouville_classical}
%%%%%%%%%%%%%%%%%%%%%%%%%%%%%%

Liouvilles theorem can be used to solve the conservative classical part of the dynamics of the Wigner function separately from the quantum dynamics, see Refs.~\cite{rodallordes2023numerical, risken1985}. The theorem states
\begin{equation}\label{eq:liouville}
    W_{\text{c}}(z,p,t)=W_{\text{c}}(z_\text{c}(z,-p,t), -p_\text{c}(z,-p,t),0),
\end{equation}
where $z_\text{c}$ and $p_\text{c}$ are the classical trajectories obtained through the Hamilton equations of motion \cite{risken1985}. In our case, these are
\begin{align}\label{eq:classical_evo}
        \nonumber \frac{\text{d}p_\text{c}}{\text{d}t}&=-2\Omega_2z_\text{c}-4\Omega_4z_\text{c}^3,\\
        \frac{\text{d}z_\text{c}}{\text{d}t}&=\frac{\wzini}{2}p_\text{c},
\end{align}
for which we use the exact solutions \cite{RieraCampeny2024analytical, salas2014} in terms of the Jacobi elliptic functions. Explicitly, the solution is given in \eref{eq:classical_solutions}.

%%%%%%%%%%%%%%%%%%%%%%%%%%%%%%
\subsection{Finding the integration constants and quadrature derivatives}\label{app:int_constants_and_derivatives}
%%%%%%%%%%%%%%%%%%%%%%%%%%%%%%

Starting from \eref{eq:classical_solutions}, the integration constants $c_1$ and $c_2$ can be found from the initial conditions by solving the equations resulting from setting $z_\text{c}(t=0)=z$ and $p_\text{c}(t=0)=p$, respectively. The resulting equations are
\begin{align}\label{eq:exact_jacobi}
        \nonumber z&=c_1\text{cn}\left(c_2, \mathfrak{m}\right)\\
        p&=-2\frac{\nu}{\omega_0} c_1\text{sn}\left(c_2, \mathfrak{m}\right)\text{dn}\left(c_2, \mathfrak{m}\right),
\end{align}
where we defined $\nu=\sqrt{\omega_0(\Omega_2+2c_1^2\Omega_4)}$ and $\mathfrak{m}=\omega_0\Omega_4c_1^2/\nu^2$. The second equation can be rewritten utilizing the relationships between the Jacobi elliptic functions to
\begin{equation}\label{eq:1st_boundary_duffing}
    p+2\frac{\nu}{\omega_0}c_1\sqrt{(1-(z/c_1)^2)[(1-\mathfrak{m})+\mathfrak{m},(z/c_1)^2]}=0
\end{equation}
which has the solution
\begin{equation}
    c_1=\pm\sqrt{\pm\sqrt{{\left(z^2+\frac{\Omega_2}{2\Omega_4}\right)}^2+\frac{\omega_0p^2}{4\Omega_4}}-\frac{\Omega_2}{2\Omega_4}},
\end{equation}
and consequently
\begin{equation}\label{eq:c_2_eval}
    c_2=\text{cn}^{-1}\left(z/c_1, \mathfrak{m}\right)=F(\arccos{(z/c_1)}, \mathfrak{m})
\end{equation}
such that both integration constants can be obtained. Here $F(\phi, \mathfrak{m})$ denotes the incomplete elliptic integral of the first kind, defined in \eref{eq:ellipF}.

The outer sign chosen for $c_1$ should be the opposite sign of $p$, and the inner sign choice should always be made positive such that $c_1\in\mathds{R}$. The reason can be seen by noticing that as a consequence of the inner sign being positive, $\nu\in\mathds{R}$ which it must be for physical reasons. With a frequency $\nu$, the solutions are doubly periodic with one real and one imaginary period such that $z_c(\nu(t+2mK(\mathfrak{m})+2niK(1-\mathfrak{m})))=(-1)^{m+n}z_c(\nu t)$ and $p_c(\nu(t+2mK(\mathfrak{m})+2niK(1-\mathfrak{m})))=(-1)^{m+n}p_c(\nu t)$ where $K(\mathfrak{m})$ is the complete elliptic integral with modulus $\mathfrak{m}$. Whenever $|\mathfrak{m}|<1$, both $K(\mathfrak{m})$ and $K(1-\mathfrak{m})$ are real. The case when $\mathfrak{m}>1$ leads to some additional insight. We first simplify the expression for the elliptic modulus, finding
\begin{equation}
    \mathfrak{m}=\frac{1}{2}-\frac{\Omega_2/4}{\sqrt{(\Omega_4z^2+\Omega_2/2)^2+\frac{\omega_0\Omega_4}{4}p^2}}.
\end{equation}
With $\Omega_2<0$, $\mathfrak{m}$ is always positive. For $\mathfrak{m}>1$, $K(\mathfrak{m})$ becomes complex, which holds when
\begin{equation}\label{eq:m_geq_1}
    \Omega_4z^4+\Omega_2z^2+\frac{\omega_0}{4}p^2 < 0.
\end{equation}
This will only take place in the DWP case and will then occur for a bounded bow-tie shaped region in phase space close to the origin. In \fref{fig:classical_dynamics}, this region is marked with gray shading. As $\text{Im}(K(\mathfrak{m}))=-K(1-\mathfrak{m})$ for $\mathfrak{m}>1$, the solutions still exhibit a real periodicity of $2K(\mathfrak{m})+2K(1-\mathfrak{m})$, but now with no sign change as both $m$ and $n$ are incremented at once. The consequence for the motion is that trajectories inside the region of $\mathfrak{m}>1$ will orbit stably in one lobe of the double well without crossing over to the other side. This phenomenon is well-illustrated in \fref{fig:classical_dynamics}(g,h,i).

Compared to the method prescribed in Ref.~\cite{rodallordes2023numerical}, utilization of the exact solutions from \eref{eq:classical_solutions} allows for direct numerical evaluation of quadrature derivatives of the type
\begin{equation}\label{eq:quad_derivative}
    \partial_p^k\bar{q}_c\equiv\frac{\partial^k q_c(z,p,-t)}{\partial p^k}\bigg\rvert_{z=z_\text{c}(z,p,t), p=p_\text{c}(z,p,t)},
\end{equation}
which are needed for performing the frame transformation. For example, with $q=z$ and $k=1$ \eref{eq:quad_derivative} corresponds to
\begin{equation}
    \partial_p \bar{z}_c=\lim_{\varepsilon\rightarrow 0}\frac{z_\text{c}(z_\text{c}(z,p,t), p_\text{c}(z,p,t)+\varepsilon, t)-z}{\varepsilon},
\end{equation}
which can now be evaluated using finite difference schemes. In the case of an arbitrary combination of uniform and hexapolar fields, the derivatives of both quadratures up to order $k=3$ with regard to $p$ needs to be evaluated at each time step in order to perform the transformation into the Liouville frame. When using second order finite differences, for each phase space discretization vertex 4 evaluations of the incomplete elliptic integral and five evaluations of the Jacobi elliptic functions \text{sn}, \text{cn} and \text{dn} are required in order to propagate $\Tilde{W}$ one time step.

%%%%%%%%%%%%%%%%%%%%%%%%%%%%%%
\subsection{Numerical evaluation of the Jacobi elliptic functions and elliptic integrals}\label{app:jacobi_eval}
%%%%%%%%%%%%%%%%%%%%%%%%%%%%%%

In solving both the case of classical and quantum dynamics, it is necessary to calculate the classical trajectories for each discrete location in phase space and simulation time. This requires evaluating a large number of Jacobi elliptic functions, some with complex arguments and some with elliptic parameter outside the standard range $\mathfrak{m}\in(0, 1)$. In order to do this, a fast converging algorithm for finding the Jacobi elliptic functions \cite{sala89}, the Jacobi imaginary transformation and the addition theorems \cite{lawden2013}, are used.
 
In order to find the integration constant $c_2$ using \eref{eq:c_2_eval}, the incomplete elliptic integral of the first kind needs to be evaluated for $\phi\in(0, \pi)$ and arbitrary real $\mathfrak{m}$. Formally the definition of this integral reads
\begin{equation}\label{eq:ellipF}
    F(\phi, \mathfrak{m}):=\int_0^{\phi}\frac{d\theta}{\sqrt{1-\mathfrak{m}\sin^2\theta}}.
\end{equation}
Fortunately, it is possible to very efficiently evaluate this integral in the full domain by using the symmetric Carlson $R_F$-function \cite{carlson1995} and the quasiperiodic extensions \cite{johansson19}. Algorithms for computing all of the special functions were implemented in Python, and JIT compiled using \texttt{numba} \cite{numba2015} to facilitate rapid evaluation.

%%%%%%%%%%%%%%%%%%%%%%%%%%%%%%
\subsection{Liouvillian operators for Wigner dynamics}\label{app:liouvillians}
%%%%%%%%%%%%%%%%%%%%%%%%%%%%%%

The quantum dynamics of the Wigner function are captured by the Liouvillian term
\begin{equation}
    \mathcal{L}_\text{q}=-8\Omega_4z\frac{\partial^3}{\partial p^3},
\end{equation}
which is found by specializing the general expression \cite{gardiner-zoller2004} to potentials up to order four in the coordinates.

In order to model the noise processes present during the time evolution, we use, as in Ref. \cite{rodallordes2023numerical}, the dissipative Liouvillian
\begin{equation}\label{eq:thermal_liouvillian}
    \mathcal{L}_\text{n}=\gamma_0\left(1+p\frac{\partial}{\partial p}\right)+\frac{2\gamma_0 k_BT}{\hbar\omega_0}\frac{\partial^2}{\partial p^2},
\end{equation}
standard to classical Liouvillian mechanics \cite{risken1985}, but rewritten in unitless quadratures. Furthermore, following Refs. \cite{rodallordes2023numerical, Romero-Isart2011}, additional momentum diffusion processes could be taken into account by adding a term
\begin{equation}
    \mathcal{L}_\text{e}=\Lambda\frac{\partial^2}{\partial p^2},
\end{equation}
which can be constructed from the term $-\frac{\Lambda}{4}\left[\hat{z}, [\hat{z}, \hat{\rho}]\right]$ in the density matrix picture by using operator correspondences to go to the phase space formulation \cite{gardiner-zoller2004}. Such a term would, for example, arise from the continuous spontaneous localization model, given that the scale of the delocalized state is much smaller than the localization length \cite{Bassi03_CSL, Abdi16}, or various noise processes including vibrational noise, magnetic field fluctuations, magnetic flux vortices, gas collisions, or black-body radiation, see, e.g.~Refs.~\cite{romero-isartLargeQuantumSuperpositions2011,PRL_Romero-Isart2012,Johnsson2016,QST_Pino2018,narasimhamoorthyMagneticNoiseMacroscopic2025,schutExpressionDecoherenceRate2025}.

%%%%%%%%%%%%%%%%%%%%%%%%%%%%%
\subsection{Solving the propagation equation for $\tilde{\mathbf{W}}(z,p,t)$}
%%%%%%%%%%%%%%%%%%%%%%%%%%%%%

Like the procedure in Ref. \cite{rodallordes2023numerical}, the differential equation resulting from \eref{eq:liouville_frame_dynamics} for the phase-space dynamics in the Liouville frame is discretized and written in the form
\begin{equation}\label{eq:discretised_prop}
    \frac{\partial \Tilde{\mathbf{W}}(t)}{\partial t}=\boldsymbol{\mathcal{D}}(t)\tilde{\mathbf{W}}(t),
\end{equation}
where $\boldsymbol{\mathcal{D}}(t)$ is a sparse matrix representation of the transformed Liouvillian operator. \eref{eq:discretised_prop} is a linear system of differential equations, and we obtained the best results both in terms of speed and accuracy by solving it using the Magnus expansion \cite{blanes2009review}. Briefly, this method entails writing the solution to \eref{eq:discretised_prop} as 
\begin{equation}
    \tilde{\mathbf{W}}(t+\text{d}t)=e^{\boldsymbol{\Omega}(t+\text{d}t,t)}\tilde{\mathbf{W}}(t).
\end{equation}
The matrix $\mathbf{\Omega}(t+\text{d}t, t)$ can then be expanded in a series $\mathbf{\Omega}(t+\text{d}t, t)=\sum_{n=1}^\infty\mathbf{\Omega}_n(t+\text{d}t, t)$, and the terms truncated at a fixed order $n$ in $\text{d}t$, where expressions for $\boldsymbol{\Omega}_n$ and efficient methods to compute them are known. We specifically adopt a 4th-order method ($\boldsymbol{\Omega}$-terms up to $n=2$) with 6th-order error control \cite{blanes_casas_ros_2000}. The procedure to take one time step starts with evaluating the $\boldsymbol{\mathcal{D}}$-matrix in order to form the necessary expansion terms for the 6th order method:
\begin{eqnarray}
    \boldsymbol{\mathcal{D}}_1&=&\boldsymbol{\mathcal{D}}\left(t+\left(\frac{1}{2}-\sqrt{\frac{3}{20}}\right)\text{d}t\right)\\
    \boldsymbol{\mathcal{D}}_2&=&\boldsymbol{\mathcal{D}}\left(t+\frac{1}{2}\text{d}t\right)\\
    \boldsymbol{\mathcal{D}}_3&=&\boldsymbol{\mathcal{D}}\left(t+\left(\frac{1}{2}+\sqrt{\frac{3}{20}}\right)\text{d}t\right)
\end{eqnarray}
from which we can calculate
\begin{eqnarray}
    \mathbf{B}^{(0)}&=&\frac{5\boldsymbol{\mathcal{D}}_1+8\boldsymbol{\mathcal{D}}_2+5\boldsymbol{\mathcal{D}}_3}{18}\\
    \mathbf{B}^{(1)}&=&\frac{\sqrt{15}}{36}(\boldsymbol{\mathcal{D}}_3-\boldsymbol{\mathcal{D}}_1)\\
    \mathbf{B}^{(2)}&=&\frac{1}{24}(\boldsymbol{\mathcal{D}}_1+\boldsymbol{\mathcal{D}}_3)
\end{eqnarray}
and then form the approximate Magnus expansion terms as
\begin{eqnarray}
    \tilde{\boldsymbol{\Omega}}_1&=&\text{d}t\mathbf{B}^{(0)}\\
    \tilde{\boldsymbol{\Omega}}_2&=&{\text{d}t}^2\left[\mathbf{B}^{(1)}, \frac{3}{2}\mathbf{B}^{(0)}-6\mathbf{B}^{(2)}\right]\\
    \nonumber \boldsymbol{V}&=&\tilde{\boldsymbol{\Omega}}_3+\tilde{\boldsymbol{\Omega}}_4\\
    \nonumber&=&\text{d}t^2\left[\mathbf{B}^{(0)}, \left[\mathbf{B}^{(0)}, \frac{1}{2}\text{d}t\mathbf{B}^{(2)}-\frac{1}{60}\tilde{\boldsymbol{\Omega}}_2\right]\right]\\
    &&+\frac{3}{5}\text{d}t\left[\mathbf{B}^{(1)}, \tilde{\boldsymbol{\Omega}}_2\right].
\end{eqnarray}
An error estimate can now be found from
\begin{equation}
    E_r=\frac{1}{2}\left|\left((\tilde{\boldsymbol{\Omega}}_1-2\mathds{1})\boldsymbol{V}-\boldsymbol{V}\tilde{\boldsymbol{\Omega}}_1\right)e^{\sum_{i=k}^4\tilde{\boldsymbol{\Omega}}_k}\tilde{\mathbf{W}}\right|,
\end{equation}
which, by standard methods can be used to propose a new timestep \cite{numerical_recipes}. Provided the error is  sufficiently small, we take then
\begin{equation}
    \tilde{\mathbf{W}}(t+\text{d}t)=e^{\sum_{k=1}^2\tilde{\boldsymbol{\Omega}}_k}\tilde{\mathbf{W}}(t)
\end{equation}
and advance the simulation by $\text{d}t$. We note that propagating one timestep with this method requires three evaluations of the $\boldsymbol{\mathcal{D}}$-matrix, as well as two calculations of a matrix exponential. For evaluating the matrix exponentials, we use the algorithm \cite{mohy-higham} as implemented in \texttt{SciPy} \cite{scipy2020}, which we modified to compute the sparse matrix operations multithreaded using MKL.

%%%%%%%%%%%%%%%%%%%%%%%%%%%%%
\subsection{Additional numerical results}\label{app:add_results}
%%%%%%%%%%%%%%%%%%%%%%%%%%%%%

%%%%%%%%%%%%%%%%%%%%%%%%%%%%%
\subsubsection{Duffing potential (DP) protocol}
%%%%%%%%%%%%%%%%%%%%%%%%%%%%%

\begin{figure*}[t!hbp]
     \centering
     \includegraphics[width=\linewidth]{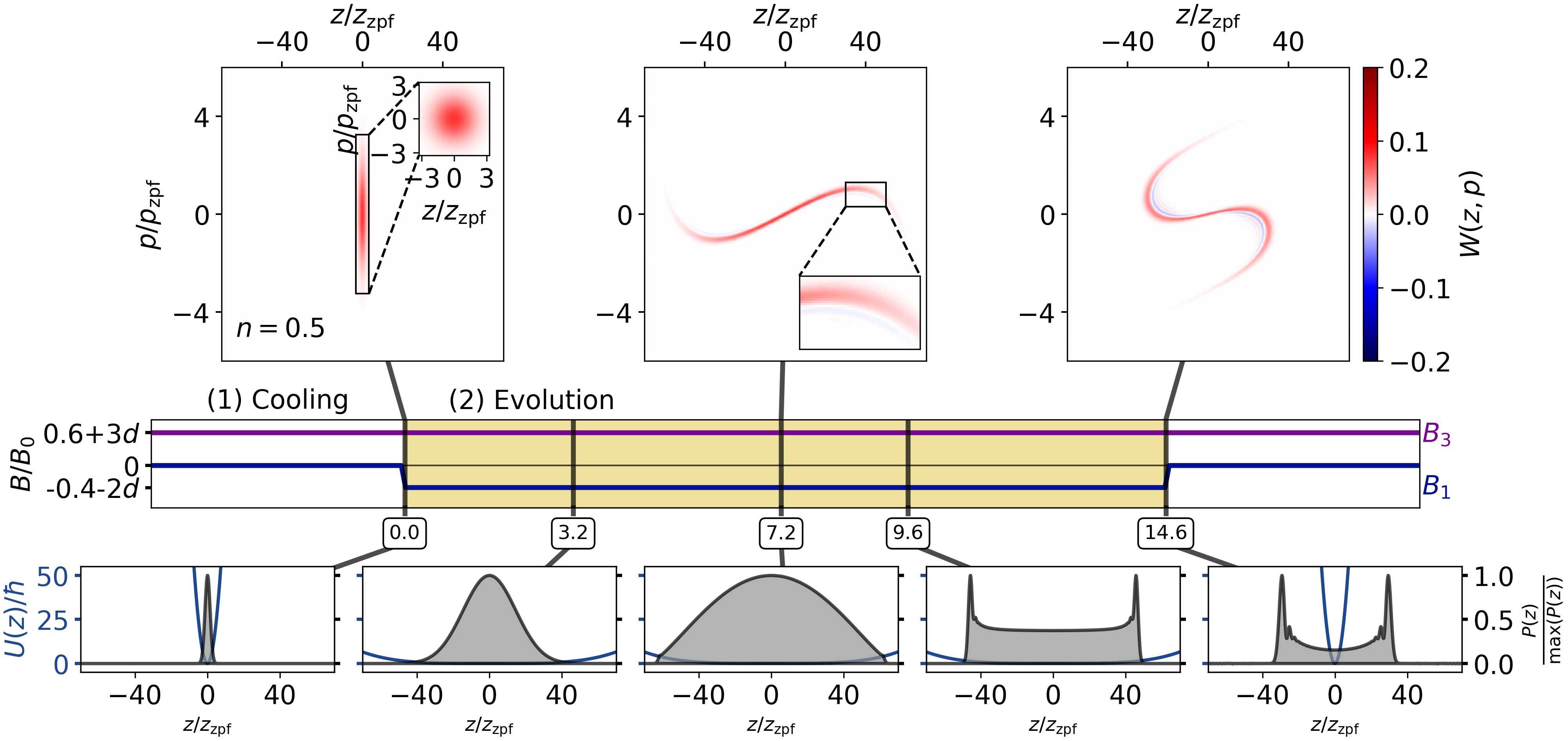}
     \caption{Non-Gaussian motional states of a levitated superconducting microparticle in the DP protocol (parameters: $\clubsuit$ in \tref{tab:parameters}, including very weak coupling to the environment with $Q=10^{12}$). Compared to \fref{fig:Wigner_protocols}, we consider here an increased initial phonon occupation of $n=0.5$. At time $t=0$, the $z^4$-potential is abruptly switched on, while maintaining $\Bh=\Bho$.  The upper panel shows the evolution of the Wigner function, the middle panel the switching of coils to realize certain magnetic fields, and the lowermost panel shows the position probability distribution $P(z)$ as well as the potential $\mathcal{U}_{z}$.}
\label{fig:Wigner_DP_App}
\end{figure*}

\fref{fig:Wigner_DP_App} shows the Wigner function evolution in the DP protocol with weak coupling to the environment, i.e., $Q=10^{12}$. The state is initially cooled in a harmonic trap to a phonon occupation $n=0.5$, higher than in \fref{fig:Wigner_protocols}. We observe that the same fringes in momentum at time $\omega_0t/2\pi=7.2$ as well as the fringes in position at time $\omega_0t/2\pi=14.6$ are reduced in amplitude compared to the case shown in \fref{fig:Wigner_protocols}.

%%%%%%%%%%%%%%%%%%%%%%%%%%%%%
\subsubsection{Double-well potential (DWP) protocol}
%%%%%%%%%%%%%%%%%%%%%%%%%%%%%

\begin{figure*}[t!hbp]
     \centering
     \includegraphics[width=\linewidth]{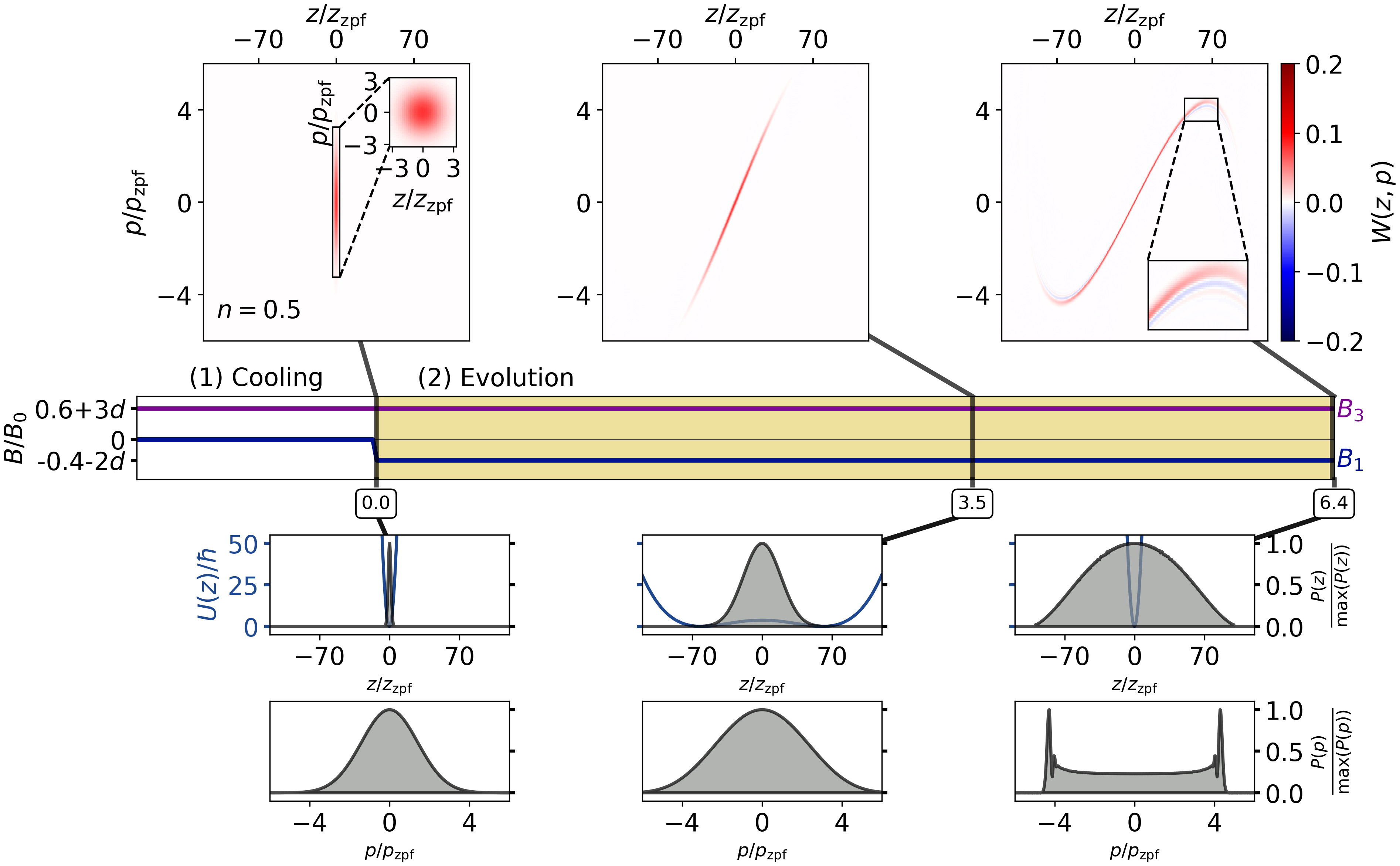}
     \caption{Non-Gaussian motional state of a levitated superconducting microparticle in the DWP protocol (parameters: $\spadesuit$ in \tref{tab:parameters}, including weak coupling to the environment with $Q=10^{12}$). At time $t=0$, the DWP-potential is abruptly switched on, while maintaining $\Bh=\Bho$ with a detuning between $\Bu$ and $\Bh$ of $d=10^{-9}$ to realize a DWP.  The upper panel shows the evolution of the Wigner function, the middle panel the switching of coils to realize the magnetic fields, and the lower two panels show the position probability distribution $P(z)$ as well as the potential $\mathcal{U}_{z}$ and the momentum probability distribution $P(p)$, respectively.}
\label{fig:Wigner_DWP_App}
\end{figure*}

\fref{fig:Wigner_DWP_App} shows the Wigner function evolution in the DWP protocol with weak coupling to the environment, i.e., $Q=10^{12}$. The state is initially cooled in a harmonic trap to a phonon occupation $n=0.5$. We observe fringes in momentum appearing at time $\omega_0t/2\pi=6.4$. Due to the emergence of numerical error, we were unable to evolve the state further in time with good accuracy. As we were able to propagate the conservative classical dynamics for virtually unlimited time, we hypothesize that this error is due to the stepping algorithm. We expect however that fringes would eventually also appear in the position probability distribution when the state is further evolved in time. We observe also that the generated state in the DWP protocol has about twice the extent in position as in the DP protocol. Therefore, we expect it to have four times the effective decay rate based on \eref{eq:decay_amplitude}. Considering limits imposed by decoherence, it may thus be advantageous to use the DP protocol over DWP.

%%%%%%%%%%%%%%%%%%%%%%%%%%%%%%
%%%%%%%%%%%%%%%%%%%%%%%%%%%%%%
\section{Parameters} \label{app:table_data}
%%%%%%%%%%%%%%%%%%%%%%%%%%%%%%
%%%%%%%%%%%%%%%%%%%%%%%%%%%%%%

\tref{tab:parameters} shows the parameters considered for the magnetic potential landscape for the different protocols.

\begin{table*}[t!hbp]
    \centering
    \caption{Rates realizable in experiments with an initial field given by $\Bqo$ or $\Bho$, depending on the protocol, assuming a Pb particle with radius $\Rp=1$~\textmu m, mass density $\varrho = 1.13\cdot10^4$~kg/m$^3$, mass $m=47$~pg and critical field $B_\text{c} = 80$~mT. Note that the initial trap frequency determines the initial field value.
    }
    \resizebox{1.0\textwidth}{!}{
        \begin{tabular}{l|c|ccccc} \hline\hline
             {Protocol} &{Field parameters} &  $\Omega_2/\omega_z^\text{ini}$ & $\Omega_4/\omega_z^\text{ini}$ & $D/z_{\text{zpf}}$ & $|\Delta E|/(\hbar \omega_z^\text{ini})$ 
             \\\hline\hline
             \multirow{4}{*}{DP-($\Bho$)} & $3\Bu=-2\Bh^\text{ini}$, $\omega_z^\text{ini}=2\pi\cdot 100$ Hz, $d=0$ &  $0$ & $3.3\cdot 10^{-13}$ & - & - & 
             \\\cline{2-7}
             & \hspace{1.8cm} $3\Bu=-2\Bh=6B_{\text{c}}/5$, $\omega_z^\text{ini}=2\pi\cdot 100$ Hz, $d=0$ \hspace{1.5cm} $\clubsuit$ &  $0$ & $2.7\cdot 10^{-7}$ & - & - & 
             \\\cline{2-7}  & $3\Bu=-2\Bh^\text{ini}$, $\omega_z^\text{ini}=\omega_{z\text{,max}}^\text{ini}(\Bho=B_\text{c})$, $d=0$ & 0 & $2.2\cdot 10^{-16}$ & - & - & 
             \\\cline{2-7}  & $3\Bu=-2\Bh=6B_{\text{c}}/5$, $\omega_z^\text{ini}=\omega_{z\text{,max}}^\text{ini}(\Bho=B_\text{c})$, $d=0$ & 0 & $7.9\cdot 10^{-17}$ & - & - & 
             \\\Xhline{1.5pt}
             %%%%%%
             \multirow{4}{*}{DWP-($\Bho$)} & $3(1-5d)\Bu=-2(1+5d)\Bh^\text{ini}$, $\omega_z^\text{ini}=2\pi\cdot 100$ Hz, $d=10^{-9}$ & $-2.5\cdot 10^{-9}$ & $3.3\cdot 10^{-13}$ & $6.1\cdot 10^1$ & $4.7\cdot 10^{-6}$ & 
             \\\cline{2-7}  & \hspace{0.5cm} $3(1-5d)\Bu=-2(1+5d)\Bh=6B_{\text{c}}/5$, $\omega_z^\text{ini}=2\pi\cdot 100$ Hz, $d=10^{-9}$ \hspace{0.2cm} $\spadesuit$ & $-2.0\cdot 10^{-3}$ & $2.7\cdot 10^{-7}$ & $6.1\cdot 10^1$ & $3.9\cdot 10^0$ & 
             \\\cline{2-7}  & $3(1-5d)\Bu=-2(1+5d)\Bh^\text{ini}$, $\omega_z^\text{ini}=\omega_{z\text{,max}}^\text{ini}(\Bho=B_\text{c})$, $d=10^{-9}$ & $-2.5\cdot 10^{-9}$ & $2.2\cdot 10^{-16}$ & $2.4\cdot 10^3$ & $7.1\cdot 10^{-3}$ & 
             \\\cline{2-7}  & $3(1-5d)\Bu=-2(1+5d)\Bh=6B_{\text{c}}/5$, $\omega_z^\text{ini}=\omega_{z\text{,max}}^\text{ini}(\Bho=B_\text{c})$, $d=10^{-9}$ & $-9.0\cdot 10^{-10}$ & $7.9\cdot 10^{-17}$ & $2.4\cdot 10^3$ & $2.6\cdot 10^{-3}$ & 
             \\\Xhline{1.5pt}
             %%%%%%
             \multirow{3}{*}{DWP-($\Bqo$)} & $3(1-5d)\Bu=-2(1+5d)\Bh=6\Bqo/5$, $\omega_z^\text{ini}=2\pi\cdot 100$ Hz, $d=10^{-9}$  & $-1.2\cdot10^{-9}$ & $1.6\cdot10^{-13}$ & $6.1\cdot10^{1}$ & $2.3\cdot 10^{-6}$ &  
             \\\cline{2-7} & $3(1-5d)\Bu=-2(1+5d)\Bh=6B_{\text{c}}/5$, $\omega_z^\text{ini}=2\pi\cdot 100$ Hz, $d=10^{-9}$ & $-2.0\cdot 10^{-3}$ & $2.7\cdot 10^{-7}$ & $6.1\cdot10^{1}$ & $3.9\cdot 10^0$ & 
             \\\cline{2-7}  & $3(1-5d)\Bu=-2(1+5d)\Bh=6\Bqo/5$, $\omega_z^\text{ini}=\omega_{z\text{,max}}^\text{ini}(\Bqo=B_\text{c})$, $d=10^{-9}$ & $-1.2\cdot 10^{-9}$ & $1.2\cdot 10^{-16}$ & $2.2\cdot 10^{3}$ & $3.0\cdot 10^{-3}$ & 
             \\\hline\hline
        \end{tabular}
    }\label{tab:parameters}
\end{table*}

%%%%%%%%%%%%%%%%%%%%%%%%%%%%%%%%%%%%%%
%%%%%%%%%%%%%%%%%%%%%%%%%%%%%%%%%%%%%%
\section{Algebraic analysis of phase space dynamics}\label{app:dynamics_ana}
%%%%%%%%%%%%%%%%%%%%%%%%%%%%%%%%%%%%%%
%%%%%%%%%%%%%%%%%%%%%%%%%%%%%%%%%%%%%%

%%%%%%%%%%%%%%%%%%%%%%%%%%%%%%%%%%%%%%
\subsection{Formal solution using Lie algebra decomposition}
%%%%%%%%%%%%%%%%%%%%%%%%%%%%%%%%%%%%%%

We derive an approximate analytical description of the open system dynamics introduced in \sref{sec:analytics}. We begin by simplifying the system Hamiltonian. Starting from \eref{eq:H_quantum_completeDP}, we first write
\begin{align}
    \hat{\mathcal{H}}/\hbar
    &= \frac{\omega_0}{4} \hat{p}^2 + \Omega_2 \hat{z}^2 + \Omega_4 \hat{z}^4 \nonumber \\
    &= \Omega_0\, \hat{a}^\dagger \hat{a} + \Omega_{\mathrm{sq}} \left(\hat{a}^{\dagger 2} + \hat{a}^2\right) + \Omega_4 (\hat{a}^\dagger + \hat{a})^4,
\end{align}
where we have dropped an irrelevant constant term and defined
\begin{align}
    \Omega_0 = \frac{\omega_0}{2} + 2\Omega_2, \qquad \Omega_{\mathrm{sq}} = -\left(\frac{\omega_0}{4} - \Omega_2\right).
\end{align}

Next, we move to a rotating frame defined by the unitary transformation $\hat{U}_0(t) = e^{-i \Omega_0 t \hat{a}^\dagger \hat{a}}$, under which the Hamiltonian becomes
\begin{eqnarray}
    \nonumber\hat{U}_0(t) (\hat{\mathcal{H}}/\hbar) \hat{U}_0^\dagger(t)
    &= &\Omega_{\mathrm{sq}} \left( e^{2 i \Omega_0 t} \hat{a}^{\dagger 2} + e^{-2 i \Omega_0 t} \hat{a}^2 \right) \\
    &&  + \Omega_4 \left( e^{i \Omega_0 t} \hat{a}^\dagger + e^{-i \Omega_0 t} \hat{a} \right)^4.
\end{eqnarray}
Applying the rotating wave approximation (RWA) in the weak quartic potential regime $\Omega_4 / \Omega_0 \ll 1$, we neglect rapidly oscillating terms in the quartic interaction and obtain 
\begin{eqnarray}
    \nonumber\hat{U}_0(t) (\hat{\mathcal{H}}/\hbar) \hat{U}_0^\dagger(t)
    &\approx& \Omega_{\mathrm{sq}} \left( e^{2 i \Omega_0 t} \hat{a}^{\dagger 2} + e^{-2 i \Omega_0 t} \hat{a}^2 \right) \\
    &&+ 6 \Omega_4 (\hat{a}^\dagger \hat{a})^2.
\end{eqnarray}
Here, we note that the squeezing terms cannot be eliminated within the RWA, as their prefactor $\Omega_{\mathrm{sq}}$ is not perturbatively small compared to $\Omega_0$. Finally, transforming back to the laboratory frame, the effective Hamiltonian takes the form \begin{align}
    \hat{\mathcal{H}}_{\mathrm{RWA}}/\hbar &= \underbrace{\Omega_0\, \hat{a}^\dagger \hat{a} + \Omega_{\mathrm{sq}} \left(\hat{a}^{\dagger 2} + \hat{a}^2\right)}_{\hat{\mathcal{H}}_{\mathrm{sq}}}+ \underbrace{\Omega_{\mathrm{Kerr}} \left(\hat{a}^\dagger \hat{a}\right)^2}_{\hat{\mathcal{H}}_{\mathrm{Kerr}}}, \label{eq_analytics_01}
\end{align}
where $\Omega_{\mathrm{Kerr}} = 6\Omega_4$, and we have grouped different dynamics according to their corresponding parameter strengths, namely the generalized squeezing $\mathcal{H}_\mathrm{sq}$, and the weak Kerr interaction $\mathcal{H}_\mathrm{Kerr}$.

Based on \eref{eq_analytics_01}, the corresponding Lindblad master equation, including single-phonon loss and thermalization, takes the form
\begin{widetext}
\begin{align}
    \frac{d}{dt}\hat{\rho}
    &= \underbrace{-i\left[\hat{\mathcal{H}}_{\mathrm{sq}}, \hat{\rho} \right]}_{\hat{\mathcal{L}}_{\mathrm{sq}}}
    + \underbrace{-i\left[\hat{\mathcal{H}}_{\mathrm{Kerr}}, \hat{\rho} \right]
    + \hat{\mathcal{D}}[\sqrt{\gamma_0 \bar{n}}\, \hat{a}^\dagger]\hat{\rho}
    + \hat{\mathcal{D}}[\sqrt{\gamma_0 (\bar{n}+1)}\, \hat{a}]\hat{\rho}}_{\hat{\mathcal{L}}_{\mathrm{Kerr}}} \nonumber \\
    &= \hat{\mathcal{L}}_{\mathrm{sq}}[\hat{\rho}] + \hat{\mathcal{L}}_{\mathrm{Kerr}}[\hat{\rho}],
    \label{eq_analytics_03}
\end{align}
\end{widetext}
where $\hat{\mathcal{D}}[\hat{O}]\hat{\rho}$ denotes the standard Lindblad dissipator
\begin{equation}\label{eq:dissipator_canonical}
    \hat{\mathcal{D}}[\hat{O}]\hat{\rho}=\hat{O} \hat{\rho} \hat{O}^\dagger-\frac{1}{2}\left\{\hat{O}^\dagger\hat{O}, \hat{\rho}\right\}
\end{equation}
for an arbitrary jump operator $\hat{O}$. Now the time evolution can be formally solved by taking the exponential of the right-hand side.

Since the dynamics generated by $\hat{\mathcal{L}}_{\mathrm{sq}}$ and $\hat{\mathcal{L}}_{\mathrm{Kerr}}$ can each be decomposed into simpler sub-dynamics using Lie-algebraic methods~\cite{chaturvedi1991solution,PRXQuantum.6.010201}, we employ the split-operator (Trotter) decomposition \cite{Gilles}
\begin{align}
    e^{(\hat{\mathcal{L}}_{\mathrm{sq}}+\hat{\mathcal{L}}_{\mathrm{Kerr}})t}
    \approx e^{\frac{1}{2}\hat{\mathcal{L}}_{\mathrm{sq}} t}
    e^{\hat{\mathcal{L}}_{\mathrm{Kerr}} t}
    e^{\frac{1}{2}\hat{\mathcal{L}}_{\mathrm{sq}} t},
    \label{eq_analytics_02}
\end{align}
which allows each exponential to be treated independently. The decomposition of $e^{\frac{1}{2}\hat{\mathcal{L}}_{\mathrm{sq}} t}$ has been obtained in \cite{PRXQuantum.6.010201} as
\begin{align}
    e^{\frac{1}{2}\hat{\mathcal{L}}_{\mathrm{sq}} t}=\hat{R}(t/2)\, \hat{\rho}\, \hat{R}^{-1}(t/2).
\end{align}
Here, $\hat{R}(t)$ is a generalized squeezing operator of the form
\begin{align}
    \hat{R}(t) = e^{-i \xi_+ \hat{A}_+} e^{-i \xi_0 \hat{A}_0} e^{-i \xi_- \hat{A}_-},
\end{align}
with generators
\begin{align}
    \hat{A}_+ = \frac{1}{2} \hat{a}^{\dagger 2}, \qquad \hat{A}_- = \frac{1}{2} \hat{a}^2, \qquad \hat{A}_0 = \frac{1}{4} \left(2 \hat{a}^\dagger \hat{a} + 1 \right),
\end{align}
and coefficients
\begin{equation}
    \begin{aligned}
        \xi_{+} = \xi_{-} & =  \frac{\Omega_{\text{sq}}}{ \Gamma}   \left[   \frac{  \sinh(\Gamma t)}{\cosh(\Gamma t) + \frac{i}{2\Gamma}  \sinh(\Gamma t) } \right] ,\\
        \xi_0 & = - 2  i \ln\left[\cosh(\Gamma t) + \frac{i}{2 \Gamma} \sinh(\Gamma t)\right],\\
        \Gamma^2 &=  \frac{\Omega_{\text{sq}}^2}{\Omega_0^2}  - \frac{1}{4}.
    \end{aligned}    
\end{equation}
The evolution generated by $\hat{\mathcal{L}}_{\mathrm{Kerr}}$ can be decomposed as \cite{chaturvedi1991solution}
\begin{align}\label{eq:skerr}
    \mathcal{S}_{\mathrm{Kerr}}(t)= e^{\hat{F}_c} e^{\hat{F}_+ \hat{K}_+} e^{\hat{F}_3 \hat{K}_3} e^{\hat{F}_- \hat{K}_-},
\end{align}
where the superoperators $\hat{K}_\bullet$ act on the density matrix as \begin{align}
    \nonumber \hat{K}_-\hat{\rho}&=\hat{a}\hat{\rho}\hat{a}^\dag,  &&\hat{K}_+\hat{\rho}=\hat{a}^\dag\hat{\rho}\hat{a},\\
    \nonumber \hat{K}_3\hat{\rho}&=\hat{a}^\dag\hat{a}\hat{\rho}+\hat{\rho}\hat{a}^\dag\hat{a},
    &&\hat{K}_0\hat{\rho}=\hat{a}^\dag\hat{a}\hat{\rho}-\hat{\rho}\hat{a}^\dag\hat{a},
\end{align}
and the coefficients $\hat{F}_\bullet$ are operator-valued functions given by
\begin{equation}
    \begin{aligned}
        \hat{F}_{+}&=\frac{2\gamma_0 \Bar{n}\tanh\left(\frac{\hat{B}}{2}t\right)}{\hat{B}+(\hat{A}+\gamma_0(2\Bar{n}+1))\tanh\left(\frac{\hat{B}}{2}t\right)},\\
        \hat{F}_{3}&=-\ln\left(\cosh\left(\frac{\hat{B}}{2}t\right)+\frac{\hat{A}+\gamma_0 (2\Bar{n}+1)}{\hat{B}}\sinh\left(\frac{\hat{B}}{2}t\right)\right),\\
        \hat{F}_{-}&=\frac{2\gamma_0 (\Bar{n}+1)\tanh\left(\frac{\hat{B}}{2}t\right)}{\hat{B}+(\hat{A}+\gamma_0(2\Bar{n}+1))\tanh\left(\frac{\hat{B}}{2}t\right)},\\
        \hat{F}_c&= \hat{F}_{3}+\frac{1}{2}\hat{A} t+\frac{\gamma_0}{2}t,
    \end{aligned}
\end{equation}
with
\begin{equation}
\begin{aligned}
    \hat{A} &= 12 i \Omega_4 \hat{K}_0, \\
    \hat{B} &= \sqrt{(\hat{A} + \gamma_0)^2 + 4 \hat{A} \gamma_0 \bar{n}},
\end{aligned}.
\end{equation}
It is convenient to work with these superoperators in the vectorized formalism, where $\hat{K}_\bullet$ are defined as
\begin{equation}
\begin{aligned}
    \hat{K}_-&=\hat{a}\otimes\hat{a},  &&\hat{K}_+=\hat{a}^\dag\otimes\hat{a}^\dag,\\
    \hat{K}_3&=\hat{a}^\dag\hat{a}\otimes\mathds{1}+\mathds{1}\otimes\hat{a}^\dag\hat{a},
    &&\hat{K}_0=\hat{a}^\dag\hat{a}\otimes\mathds{1}-\mathds{1}\otimes\hat{a}^\dag\hat{a}.
\end{aligned}
\end{equation}
from which their action on the Fock basis vectorized density matrix can now be understood.

%%%%%%%%%%%%%%%%%%%%%%%%%%%%%%%%%%%%%%
\subsection{Equivalence of the environmental couplings}\label{app:lindbladian_equivalence}
%%%%%%%%%%%%%%%%%%%%%%%%%%%%%%%%%%%%%%

To illustrate how the environmental couplings in the phase space and density matrix picture are connected, we separately convert the dissipator terms into the phase space picture according to the correspondence rules established in \cite{gardiner-zoller2004}. The dissipator term is given in \eref{eq:dissipator_canonical}. In the case of the terms in \eref{eq_analytics_03}, using the correspondence rules yields
\begin{widetext}
\begin{equation}\label{eq:density_thermal_coupling}
    \hat{\mathcal{D}}[\sqrt{\gamma_0\bar{n}}\hat{a}^\dagger]\hat{\rho}+\hat{\mathcal{D}}[\sqrt{\gamma_0(\bar{n}+1)}\hat{a}]\hat{\rho} \leftrightarrow \frac{\gamma_0}{2}\left[2+z\frac{\partial}{\partial z}+p\frac{\partial}{\partial p}+(2\Bar{n}+1)\left(\frac{\partial^2}{\partial z^2}+\frac{\partial^2}{\partial p^2}\right)\right]W(z,p,t).
\end{equation}
\end{widetext}
The right hand side expression for the Wigner function environmental coupling term is different from the momentum diffusion process modeled in \eref{eq:thermal_liouvillian} used in the phase space numerical analysis, as it distributes the diffusion equally into momentum and position diffusion. However, for small frequency oscillators at high environmental temperature ($k_BT \gg \hbar\omega_0$), the two different noise models are expected to be equivalent \cite{Ghosh23}. We thus posit that \eref{eq:density_thermal_coupling} can be used to model the environmental coupling $\mathcal{L}_{\text{e}}$ in the density matrix picture, despite the correspondence not being exact.

%%%%%%%%%%%%%%%%%%%%%%%%%%%%%%%%%%%%%%
\subsection{Scaling relations}\label{app:scalingrelations}
%%%%%%%%%%%%%%%%%%%%%%%%%%%%%%%%%%%%%%

We show now how we obtain the scaling relations that are used to fit the data of  \fref{fig:feasibility}(d,f,h). However, we can currently not provide a scaling relation for the data of \fref{fig:feasibility}(b). The latter would require a combination of the amplitude reduction due to phonon loss and of a broadened phonon distribution due to thermalization, which we leave to future work. 

%%%%%%%%%%%%%%%%%%%%%%%%%%%%%%%%%%%%%%
\subsubsection{Fringe amplitude}
%%%%%%%%%%%%%%%%%%%%%%%%%%%%%%%%%%%%%%

The term $e^{\hat{F}_3 \hat{K}_3}$ from \eref{eq:skerr} governs the amplitude of the density matrix elements. In particular, it determines the decay of the amplitude of the interference fringes in phase space. An associated effective decay factor $\gamma_{\mathrm{eff}} t$ can be estimated via the truncated operator norm \cite{Winter17,Shirokov_2020,Becker25} as
\begin{align}\label{eq:decay_amplitude2}
    \gamma_{\mathrm{eff}} t&\equiv \| \hat{F}_3 \hat{K}_3 \|_{z_0} \nonumber \\
    &= \max_{m,n \in \{0,1,\dots,n_0\}} \langle m,n | \hat{F}_3 \hat{K}_3 | m,n \rangle \nonumber \\
    &= \left| -2n_0\ln((\bar{n}+1)e^{\gamma_0 t/2}-\bar{n}e^{-\gamma_0 t/2}) \right| \nonumber \\
    &\approx {z_0}^2\gamma_0 t (1 + 2\bar{n})/4,
\end{align}
where in the last line we assumed $\gamma_0 t\ll 1$, the truncation was performed at Fock state $|n_0\rangle$, with $n_0 \approx z_0^2/4$, and $z_0$ denotes the phase-space coordinate at which the fringe amplitude is evaluated. Here, the vectorized formalism has been used, in which the density matrix entry $|m\rangle\langle n|$ becomes $|m\rangle\otimes|n\rangle\equiv |m,n\rangle$. 

Although in the presence of Kerr nonlinearity the superoperator $\hat{F}_3 \hat{K}_3$ can acquire complex values when acting on off-diagonal density matrix elements, where the real part corresponds to their contribution to decay and the imaginary part to nonlinear phase evolution, the maximal decay rate is always determined by the diagonal elements. Visually, this can be understood from the aspect that the quantum Kerr interaction primarily adds a phase to the classical phase space structure, which in turn induces an oscillating negative fringe pattern. That is, the decay of these fringes is still ultimately governed by the decay of the diagonal population that supports the pre-existing positive pattern. This population is determined by the action of the squeezing operation on the initial state $e^{\frac{1}{2}\hat{\mathcal{L}}_\mathrm{sq} t}$ according to \eref{eq_analytics_02}. 

%%%%%%%%%%%%%%%%%%%%%%%%%%%%%%%%%%%%%%
\subsubsection{Fringe wavenumber}\label{app:fringewavevector}
%%%%%%%%%%%%%%%%%%%%%%%%%%%%%%%%%%%%%%

The Wigner function expressed in polar coordinates $r = \sqrt{z^2 + p^2}$ and $\phi = \tan^{-1}(p/z)$ for a density matrix $\rho_0 = \sum_{m,n} c_{m,n} |m\rangle\langle n|$ evolving under the Kerr interaction in a closed system can be written as \begin{equation}\label{eq:wignerkerr}
    W(r,\phi)=\sum_{m=0}^\infty \sum_{k=-m}^\infty c_{m,m+k}f_{m,k}(r,\phi),
\end{equation}
with
\begin{align}
    f_{m,k}(r,\phi)&=\begin{cases}
        \sqrt{\frac{m!}{(m+k)!}} e^{-ik(\phi-(2m+k)\mathcal{K}_2t_0)}\times\nonumber\\
    \hspace{0.15cm}\frac{(-1)^m}{\pi}(2r^2)^{\frac{k}{2}}L_m^{k}(2r^2)e^{-r^2} \hspace{1.03cm} \text{if}\ k\geq 0,\\
    \sqrt{\frac{(m+k)!}{m!}} e^{-ik(\phi-(2m+k)\mathcal{K}_2t_0)}\times\nonumber\\
    \hspace{0.15cm}\frac{(-1)^{m+k}}{\pi}(2r^2)^{-\frac{k}{2}}L_{m+k}^{-k}(2r^2)e^{-r^2} \hspace{0.29cm} \text{if}\ k< 0,
    \end{cases}
    \label{eq_wigner_polar_coord}
\end{align}
where $L_m^n(\bullet)$ is the associated Laguerre polynomial \cite{L_pez_Carre_o_2025, curtright_fairlie_zachos_2013}. 

Crucially, the index $k$ quantifies the amount of off-diagonality of the density matrix element $c_{m,m+k}$, and thereby directly controls the angular oscillation of the Wigner function. The latter can in particular be seen via the term proportional to $e^{-i k (\phi - ...)}$. We can, thus, assume that the index $k$ is proportional to a wavenumber. This also means that larger values of $k$ correspond to faster oscillations in the $\phi$ direction, which are characterized by a larger wavenumber $k_\text{max}$. Since the index $k$ is bounded from below $-m\leq k< \infty$, and we are interested in the large $|k|$ regime, we may focus only on the positive part $k\geq 0$. 

A naive expectation would suggest that enhancing populations at higher Fock states should increase the dominant wavenumber $k_\text{max}$. However, as observed in \fref{fig:feasibility}(d), more strongly thermalized states instead exhibit a peak at lower wavenumber. This apparent discrepancy arises because the Wigner function involves a summation over $m$, and the dominant prefactor at large $k$ is (using Stirlings formula)
\begin{align}
    \frac{1}{\sqrt{(m+k)!}} & \sim (m+k)^{-k/2},
\end{align}
which suppresses contributions from large $k$. 

We stress that while there are different functions involving $k$ within the Wigner function of \eref{eq:wignerkerr}, such as $e^{k^2}$, $a^k$, and $(m+k)!$, the form that increases the fastest with an increasing $k$ is the function $(m+k)^k$ coming from applying the Stirling formula on the factorial term $(m+k)!$. The other term in the Stirling formula $(m+k)^m$ grows much slower, so we ignore it, as including it would contribute merely another parameter $D$ as $e^{\frac{1}{k+D}}$ in \eref{eq_scaling_behavior}. We also note that exponential factors of the form $\sim e^{-k}$ can be effectively absorbed into fitting prefactors associated with $m$ and $k$, such that the dominant scaling of the factorial term is governed by the functional dependence $k^k$. Consequently, rapidly oscillating contributions carry exponentially small weight and can be neglected within the RWA. 

The dominant contribution therefore arises from an intermediate value $k_0$, which is smaller than the expectation based solely on the wavenumber. This value is set by a competition between wavenumber and amplitude suppression: higher $k$ contributions are suppressed by the RWA, while terms with appreciable weight are typically associated with smaller $k$. 

Based on this balance, we estimate the relevant threshold via $(m+k)^{-k/2} \sim \epsilon$, where $\epsilon$ denotes a rough compromise above which contributions are significant but do not correspond to a maximum wavenumber, and below which the wavenumber becomes sufficiently high but their contributions are negligible. We can rewrite this relation as
\begin{align}
    m+k &\sim \epsilon^{-2/k}.
\end{align}
To arrive at a scaling relation that we can use for fitting, we write $m\sim 1/A\cdot n$ with the mean phonon occupation $n$ of the oscillator. This relation follows from the fact that the  mean phonon occupation and the Fock state phonon number both quantify the number of phonons in the system and, thus, must follow a linear relation. We furthermore relate $k$ linearly with the maximal wavenumber as $k\sim C/A\cdot k_\mathrm{max}$. The latter follows from our argument that $k$ quantifies the rate of change of the fringes, which is quantified by the wavenumber. We further define $\epsilon^{-2A/C}=e^{B}$ for simplicity, and obtain
\begin{align}
    n &\sim A e^{\frac{B}{k_\mathrm{max}}} - C k_\mathrm{max},
    \label{eq_scaling_behavior}
\end{align}
with $A$, $B$, and $C$ are fitting parameters characterizing the scaling behavior. We used \eref{eq_scaling_behavior} to fit to our numerical results, the result is shown in \fref{fig:feasibility}(d).

To arrive at a scaling relation between $Q$ and $k_\mathrm{max}$, we make the assumption that the dynamics of a thermal state in a closed system [\fref{fig:feasibility}(c,d)] is equivalent to the dynamics of a vacuum state in a hot thermal environment  [\fref{fig:feasibility}(g,h)], i.e., when $k_BT\gg \hbar\omega_0$. In both cases, the Fock state distribution is essentially broadened by populating higher phonon number states, thereby increasing the mean phonon occupation. Since this redistribution plays an analogous role in both cases, we treat them on equal footing. Thus, for the case of an initial vacuum state evolving in a thermal environment, the relevant occupation number can be approximated by the time-dependent mean thermal population
\begin{align}
    n \;\to\; \bar{n}\left(1 - e^{-\gamma_0 t}\right) 
    \;=\; \bar{n}\left(1 - e^{-\frac{\omega_0 t}{Q}}\right),
\end{align}
leading to the modified scaling relation
\begin{align}
    \bar{n}\left(1 - e^{-\frac{\omega_0 t}{Q}}\right)
    &\sim A' e^{\frac{B'}{k_\mathrm{max}}} - C' k_\mathrm{max},
\end{align}
which can be written as
\begin{align}\label{eq:Qfit}
    -\frac{\omega_0 t}{Q}
    &\sim \ln\left(1 - \frac{A'}{\bar{n}} e^{\frac{B'}{k}} + \frac{C'}{\bar{n}} k \right).
\end{align}
A fit to the numerical results using \eref{eq:Qfit} is shown in \fref{fig:feasibility}(h).

\bibliography{Bib_HOT_traps}

\end{document}